\documentclass{article}

\usepackage{arxiv}

\usepackage[utf8]{inputenc} 
\usepackage[T1]{fontenc}    
\usepackage{hyperref}       
\usepackage{url}            
\usepackage{booktabs}       
\usepackage{amsfonts}       
\usepackage{amsmath}
\usepackage{amssymb}
\usepackage{nicefrac}       
\usepackage{microtype}      
\usepackage{cleveref}       
\usepackage{lipsum}         
\usepackage{graphicx}
\usepackage{natbib}
\usepackage{doi}
\usepackage{authblk}

\usepackage{overpic}
\usepackage{lmodern} 
\usepackage{xspace} 
\usepackage{subcaption}

\newcommand{\sd}{\, {\rm d}}

\usepackage{tikz}
\usetikzlibrary{shapes}

\title{Periodically activated physics-informed neural networks for assimilation tasks for three-dimensional Rayleigh-B\'enard convection}

\newbox{\orcid}\sbox{\orcid}{\includegraphics[scale=0.06]{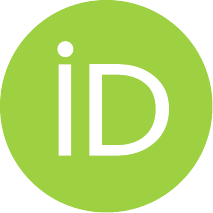}} 
\author[1]{%
	\href{https://orcid.org/0000-0002-7817-3388}{\usebox{\orcid}\hspace{1mm}Michael~Mommert}\thanks{\texttt{michael.mommert@dlr.de}}%
}
\author[1,2]{%
	Robin~Barta%
}
\author[1]{%
	Christian~Bauer%
}
\author[1]{%
	Marie-Christine~Volk%
}
\author[1,2]{%
	Claus~Wagner%
}
\affil[1]{Department Ground Vehicles, Institute of Aerodynamics and Flow Technology, German Aerospace Center, Bunsenstr. 10, 37073 Göttingen, Germany}
\affil[2]{Institute of Thermodynamics and Fluid Mechanics, Technische Universität Ilmenau, 98684 Ilmenau, Germany}

\renewcommand{\shorttitle}{Assimilation for 3D Rayleigh-B\'enard convection with PINNs}

\hypersetup{
pdftitle={Periodically activated physics-informed neural networks for assimilation tasks for three-dimensional Rayleigh-B\'enard convection},
pdfsubject={},
pdfauthor={M. Mommert, R. Barta, C. Bauer, M.-C. Volk, C. Wagner},
pdfkeywords={Rayleigh-B\'enard convection, physics-informed neural networks, assimilation, machine learning, activation functions},
}

\begin{document}
\maketitle

\begin{abstract}
We apply physics-informed neural networks to three-dimensional Rayleigh-B\'enard convection in a cubic cell with a Rayleigh number of $\mathrm{Ra}=10^6$ and a Prandtl number of $\mathrm{Pr}=0.7$ to assimilate the velocity vector field from given temperature fields and vice versa. With the respective ground truth data provided by a direct numerical simulation, we are able to evaluate the performance of the different activation functions applied (sine, hyperbolic tangent and exponential linear unit) and different numbers of neurons ($32$, $64$, $128$, $256$) for each of the five hidden layers of the multi-layer perceptron.
The main result is that the use of a periodic activation function (sine) typically benefits the assimilation performance in terms of the analyzed metrics, correlation with the ground truth and mean average error. 
The higher quality of results from sine-activated physics-informed neural networks is also manifested in the probability density function and power spectra of the inferred velocity or temperature fields.
Regarding the two assimilation directions, the assimilation of temperature fields based on velocities appears to be more challenging in the sense that it exhibits a sharper limit on the number of neurons below which viable assimilation results can not be achieved.
\end{abstract}


\section{Introduction}

Turbulent thermal convection governs a wide variety of flows considered for solving engineering tasks (see e.g. \cite{Schmeling2020}) and environmental issues (see e.g. \cite{Voelker2002}).
Understanding and controlling these flows requires knowledge of the heat fluxes they induce.
The acquisition of heat fluxes requires both velocity and temperature data.
In terms of measurements, the simultaneous recording of both fields has been shown to be feasible in a laboratory environment (for recent examples see \cite{Mommert2023, Kaeufer2023a}). However, even on this scale, these measurements require a high level of experimental effort. 

In this context, the approach of assimilating temperature fields from volumetric velocity measurements is an attractive option to avoid the significant increase in complexity that an additional temperature measurement represents.
More specifically, the assimilation based on physics-informed neural networks (PINNs) is controlled by the underlying partial differential equations (PDEs) as part of the loss function for a network, and thus does not require additional data to train the models a priori.
This approach also promises to be extensible to flows of greater complexity and scale beyond the laboratory by exploiting the potential of modern, highly parallelized computing architectures (\cite{Vinuesa2023}).
Therefore, we pursue the development of a PINN-based assimilation method that relies only on field data typically obtained from measurements (either just temperatures or just velocities).

The following are examples of applications of PINNs that are relevant to this application:
One of the possible applications of PINNs is similar to that of a numerical solver,  where the only data provided are the initial and boundary conditions.
Such implementations are usually confined to two dimensions, such as the vortex shedding behind a circular object underlying the Navier-Stokes equations, as presented in the original PINN paper by \cite{Raissi2019}. Other examples are neural solvers for the Poisson equation (\cite{Markidis2021a}) or the compressible Euler equations (\cite{Wassing2024}).
In summary, these examples show that neural networks are able to approximate the solutions of the PDEs governing the respective flows by minimizing the losses representing the residuals of the PDEs.

Yet, a major advantage of PINNs is their ability to incorporate any kind of additional data that helps to find the desired solution more effectively, as is the case for the assimilation tasks at hand (\cite{Cai2021a}).
An example of this is the use of a PINN to regularize the results of two-dimensional three-component particle image velocimetry data, shown by \cite{Hasanuzzaman2023}.
Furthermore, \cite{ClarkDiLeoni2023b} used velocity data from three-dimensional Lagrangian measurements to assimilate the pressure.
The reconstruction of velocity and pressure fields was also achieved by \cite{Eusebi2024}, showing the reconstruction of the flow of a tropical cyclone from spatially sparse data.
Another example is provided by \cite{MorenoSoto2024}, where the PINN regularizes the velocity fields of a preceding reconstruction of small time scales and provides additional pressure data.
This set of examples demonstrates the basic ability of the PINN method to handle noisy or sparse data, which is required for assimilation tasks.

The case of Rayleigh-B\'enard convection considered here is characterized by the transport of the scalar temperature.
How data of a transported scalar can be leveraged to assimilate the velocity field was shown by \cite{Raissi2020} for the three-dimensional flow in an intracranial aneurysm.
Even more relevant for Rayleigh-B\'enard convection, the temperature field was used by \cite{Cai2021} to reconstruct three-dimensional velocity and pressure fields of the region above an espresso cup in which a tomographic background-oriented Schlieren measurement was applied.
The same type of assimilation to provide temperature data for the training was done by \cite{Lucor2022}. However, they still provided both velocities and temperatures as initial conditions, which would not be feasible in the context of avoiding the respective measurement effort.
Another example of the assimilation of velocities is the work of \cite{ClarkDiLeoni2023a}, who investigated the effect of sparsity on the provided temperature data in two-dimensional Rayleigh-B\'enard convection.
\cite{Cai2021b} also used local temperature measurements in combination with boundary conditions to recover the flow and thermal boundary conditions for either forced or mixed convective flows past cylinders.

The above-mentioned publications show that PINNs are usually used to assimilate flows based on temperature data, while modern measurement techniques are capable of providing highly resolved velocity data of thermally-driven convective flows.
Therefore, we use the assimilation of velocity fields as a link to previous studies before investigating the temperature assimilation pursued, which further allows a comparison of both assimilation directions.

Also, a large part of the studies mentioned above deal with a limited amount of turbulence-induced complexity, as the variety of structure sizes also poses high requirements on PINNs to be able to reconstruct them.
This issue is, for example, addressed by implementing periodic activation functions, typically sine (\cite{Sitzmann2020}). Because of their relation to Fourier features (\cite{Tancik2020}) and thus to the idea of decomposing the flow fields into Fourier modes, they seem predestined to allow neural networks to efficiently map complex flows.
In the context of turbulent flows, periodically activated PINNs have been used by \cite{Angriman2023} to reconstruct turbulence from underresolved flow fields by imposing information about higher order statistical moments.

To evaluate the performance of the PINNs, we base our investigation on ground truth data generated in a direct numerical simulation described in section~\ref{sec:DNS}.
Regarding turbulent Rayleigh-B\'enard convection, the assimilation of velocity or temperature data becomes increasingly complex with increasing turbulence intensity, indicated by the dimensionless Rayleigh ($\mathrm{Ra}$) and Prandtl ($\mathrm{Pr}$) numbers. Therefore, we chose to study a case of Rayleigh-B\'enard convection in a cubic cell at $\mathrm{Ra}=10^6$ and $\mathrm{Pr}=0.7$ with moderate turbulence.
Since we investigate the influence of different activation functions on the ability of a neural network to map the complexity of turbulent flows, we also vary the number of neurons in each hidden layer of the PINN. Further details of the PINN and its training parameters are described in section~\ref{sec:ML}.
The results of the PINNs with the different parameters and for both assimilation directions are then presented and discussed in section~\ref{sec:results}.

\section{Generation of ground truth data}
\label{sec:DNS}
\begin{figure}[!ht]
	\centering
	\includegraphics[width=0.5\linewidth]{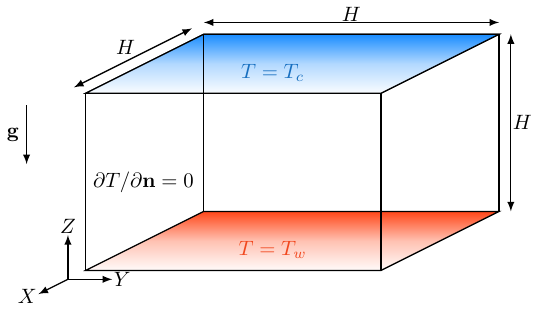}
	\caption{Cubic Rayleigh-B\'enard convection cell with the height $H$ and the volume $H^3$. All walls are no-slip boundaries; top and bottom walls are iso-thermal and side walls are adiabatic.}
	\label{fig:geometry}
\end{figure}

The ground truth data are generated in a direct numerical simulation (DNS) of a turbulent Rayleigh-B\'enard convection at $\mathrm{Pr}=0.7$ and $\mathrm{Ra}=10^{6}$ in a cubic domain, solving the transport equations for mass, momentum, and temperature for an incompressible fluid and the Boussinesq approximation,
\begin{eqnarray}
	\frac{\partial \mathbf{u}}{\partial t} + \mathbf{u} \cdot \nabla \mathbf{u} & = & -\nabla p +\sqrt{\mathrm{Pr}/\mathrm{Ra}}\nabla^2 \mathbf{u} + T \mathbf{e}_z,\label{eq:momentum}\\
	\frac{\partial T}{\partial t} + \mathbf{u} \cdot \nabla T & = & \sqrt{1/(\mathrm{Pr}\mathrm{Ra})}\nabla^2 T,\label{eq:temperature} \\
	\nabla \cdot \mathbf{u} & = & 0,\label{eq:mass}
\end{eqnarray}
where $\textbf{u}=(u_x,u_y,u_z)$ is the velocity vector, $p$ is the pressure, $T$ is the temperature, and $\mathbf{e}_z$ is the unit vector with respect to the vertical direction.
The flow geometry, consisting of a cubic domain with a heated bottom plate ($T=T_w$), a cooled top plate ($T=T_c$),
and adiabatic side walls, is displayed in Fig.~\ref{fig:geometry}.

Velocities are non-dimensionalized with the free-fall velocity $\hat{u}_\mathrm{ff}=(\hat{\alpha}\hat{g}\Delta \hat{T}\hat{H}^3)^{1/2}$, spatial coordinates with the cell height $\hat{x}_\mathrm{ref}=\hat{H}$, the time coordinate with the corresponding reference time $\hat{t}_\mathrm{ff}=\hat{x}_\mathrm{ref}/\hat{u}_\mathrm{ff}$, and the pressure with the reference pressure $\hat{p}_\mathrm{ref}=\hat{\rho}\hat{u}^2_\mathrm{ff}$ with $\hat{\rho}$ being the fluid density, $\hat{\alpha}$ the thermal diffusivity, and $\hat{g}$ the gravitational acceleration.
The temperature is non-dimensionalized by $T=(\hat{T}-\hat{T}_0)/\Delta \hat{T}$ with $\Delta \hat{T}=\hat{T}_w-\hat{T}_c$ and $\hat{T}_0=(\hat{T}_w-\hat{T}_c)/2$.
No-slip and impermeability boundary conditions are applied to all walls.
In addition, the top and bottom plates are assumed to be iso-thermal, whereas the side walls are modeled as adiabatic.
After non-dimensionalization, the equations~(\ref{eq:momentum})-(\ref{eq:mass}) are discretized spatially and temporally with a fourth-order accurate finite volume scheme and a second-order accurate Euler-leapfrog time integration scheme, respectively.
According to \cite{Wagner1994}, the temporal discretization of equation (\ref{eq:momentum}) for a leapfrog time step yields
\begin{equation}
	\frac{1}{2\Delta t}(\mathbf{u}^{n+1}-\mathbf{u}^{n-1})+\mathbf{u}^n\cdot\nabla\mathbf{u}^n  = -\nabla p^n  + \sqrt{\mathrm{Pr}/\mathrm{Ra}} \nabla^2 \mathbf{u}^{n-1} + T^n\mathbf{e}_z, \label{eq:momdisc}
\end{equation}
where $n$ is the number of the time step, $\Delta t$ is the temporal increment between two time steps, and $\delta^j_i$ is the Kronecker delta.
To integrate the equation (\ref{eq:momdisc}) in time, a fractional step algorithm as the one introduced by \cite{Chorin1967,Chorin1968} is applied.
In a first step, an auxiliary velocity vector field $\textbf{u}^*$ is estimated from equation (\ref{eq:momdisc}) neglecting the pressure term,
\begin{equation}
	\frac{1}{2\Delta t}(\mathbf{u}^{*}-\mathbf{u}^{n-1})+\mathbf{u}^n\cdot\nabla\mathbf{u}^n = \sqrt{\mathrm{Pr}/\mathrm{Ra}} \nabla^2 \mathbf{u}^{n-1} + T^n \mathbf{e}_z. \label{eq:uaux}
\end{equation}
In the second step, the pressure Poisson equation is solved for the auxiliary field as follows:
\begin{equation}
	\nabla^2 \phi^n = \nabla\cdot \mathbf{u}^*, \label{eq:poisson}
\end{equation}
with $\phi^n = 2\Delta t p^n$ and the boundary condition ($\mathbf{n}\cdot\nabla \phi^n)|_{\partial \gamma}=0$.
Due to the inhomogeneity of the flow problem with respect to all three directions, the Poisson solver utilizes a separation of variables method as described by \cite{Shishkina2009}.
Finally, the velocity at time step $n+1$ is corrected via
\begin{equation}
	\mathbf{u}^{n+1} = \mathbf{u}^* - \nabla \phi^n. \label{eq:velupdate}
\end{equation}
Regarding the spatial discretisation, the numerical grid aims to fully resolve the smallest velocity and temperature scales, which are the Kolmogorov and Batchelor length scales
\begin{eqnarray}
	\eta_K & = & \left(\frac{\nu^3}{\varepsilon_u}\right)\ \mathrm{and}\label{eq:kolmogorov}\\
	\eta_B & = & \left(\frac{\nu\kappa^2}{\varepsilon_u}\right)=\eta_K\mathrm{Pr}^{-1/2},\label{eq:batchelor}
\end{eqnarray}
respectively, with $\varepsilon_u=\nu \langle \partial u_i/\partial x_j \rangle$ the kinetic dissipation rate.
Angle brackets indicate statistical averaging.
For the present case of $\mathrm{Pr}<1$, the Kolmogorov length scale is smaller than the Batchelor length scale, and thus, more restrictive with respect to the grid resolution.
According to \cite{Shishkina2010}, the minimum grid spacing in the bulk flow region can be estimated via the following approximation of the global Kolmogorov scale
\begin{equation}
	h^{bulk}/H\le\frac{\mathrm{Pr}^{1/2}}{((\mathrm{Nu}-1)\mathrm{Ra})^{1/4}} \text{, for Pr}<1,\label{eq:esthbulk}
\end{equation}
whereas the minimum grid spacing in the thermal and viscous boundary layers is estimated based on the Prandtl-Blasius boundary layer theory,
\begin{equation}
	h^{BL}/H \le 2^{-3/2}a^{-1}\mathrm{Nu}^{-3/2}\mathrm{Pr}^{0.5355-0.033\log\mathrm{Pr}}\text{, for }3\times10^{-4} \le \mathrm{Pr}\le 1,\label{eq:esthbl}
\end{equation}
with $a\approx0.922$~\citep{Stevens2013}.
Moreover, \cite{Shishkina2010} recommend minimum numbers of nodes in the thermal and kinetic boundary layer of
\begin{eqnarray}
	N^{min}_{\delta_\theta} & = & \sqrt{2}a\mathrm{Nu}^{1/2}\mathrm{Pr}^{-0.5355+0.033\log\mathrm{Pr}} \text{, for }3\times10^{-4}\le \mathrm{Pr}\le 1,\label{eq:minnodesth}\\
	N^{min}_{\delta_u}      & = & \sqrt{2}a\mathrm{Nu}^{1/2}\mathrm{Pr}^{-0.1785+0.011\log\mathrm{Pr}} \text{, for }3\times10^{-4}\le \mathrm{Pr}\le 1.\label{eq:minnodesu}
\end{eqnarray}
With the Nusselt number derived a priori from the Grossmann-Lohse theory~\citep{Grossmann2000,Grossmann2001,Grossmann2002,Grossmann2004}, a minimum grid spacing of 0.04 in the bulk flow region and 0.02 in the boundary layer is obtained.
Additionally, the minimum number of nodes in the boundary layers \textemdash\ equations (\ref{eq:minnodesth}) and (\ref{eq:minnodesu}) \textemdash\ is estimated to $N^{min}_{\delta_\theta}=3$ and $N^{min}_{\delta_u}=3$ for the thermal and kinetic boundary layer, respectively.
Overall, an equidistant grid spacing of $\Delta z/H=0.0156$ in all three dimensions is applied (see table~\ref{tab:case}).

After an initial transient, when the average Nusselt number computed from the cooled and the heated plate,
\begin{equation}
	\mathrm{Nu}=-\frac{1}{2}\left( \left\langle \frac{\sd T }{\sd z}\right\rangle_{xyt}(z=0)+ \left\langle \frac{\sd T}{\sd z} \right\rangle_{xyt}(z=H)\right)\label{eq:nusselt},
\end{equation}
has reached a statistically stationary state, instantaneous flow field realizations in the form of velocity, pressure, and temperature fields are written out every 0.05 dimensionless time units.
The velocity fields are then interpolated from the staggered grid used in the DNS solver to a collocated grid with the pressure and temperature fields.
Hereafter, these fields (denoted with $\tilde{\cdot}$) serve as the ground truth and input data for the PINN.
\begin{table}[!bth]
	\centering
	\caption{Rayleigh-B\'enard convection simulation case. Ra is the Rayleigh, Pr the Prandtl number. $N_x$, $N_y$ and $N_z$ are the number of grid points in $x$, $y$, and $z$ direction, respectively. $\Delta z$ is the grid spacing. $N_{\delta_\theta}$ is the number of grid points in the thermal, $N_{\delta_u}$ in the kinetic boundary layer. The Nusselt number (see eq.~\ref{eq:nusselt}) is computed a posteriori.}
	\begin{tabular}{lcccccc}
		\hline\hline
		Ra & Pr & $N_x\times N_y \times N_z$ & $\Delta z/H$  & $N_{\delta_\theta}$ & $N_{\delta_u}$ & Nu  \\
		$10^{6}$ & $0.7$ & $64\times64\times64$ & $0.0156$ &  3 &  3   & 8.3    \\
		\hline\hline
	\end{tabular}
	\label{tab:case}
\end{table}

\section{PINN and training configuration}
\label{sec:ML}

The following subsections are dedicated to the PINN approach and its application to assimilation. It covers the construction of the neural network \ref{sec:ML_arch}, the data input \ref{sec:ML_data} and additional details needed to configure the training process \ref{sec:ML_opt}.

\subsection{Architecture}
\label{sec:ML_arch}

\begin{figure}[!ht]
	\centering
	\includegraphics[width=0.99\linewidth]{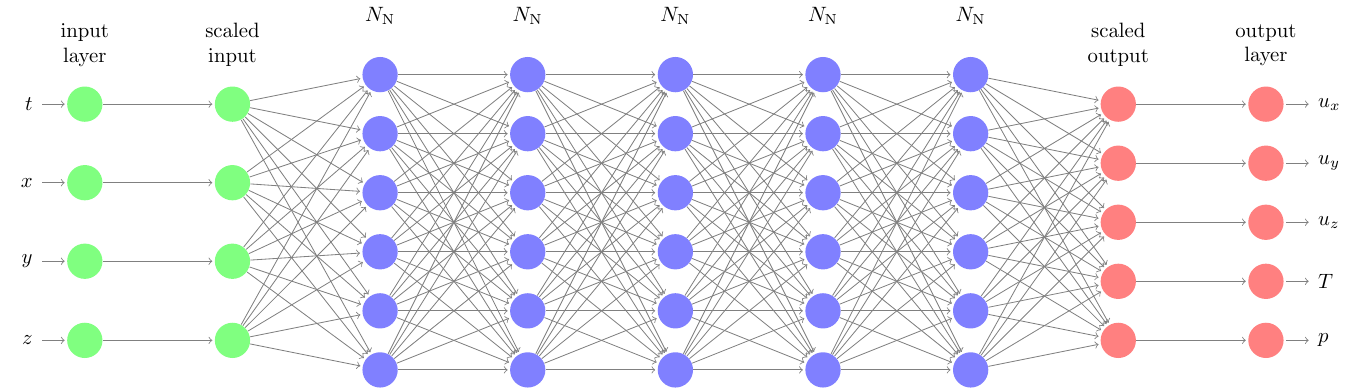}
	\caption{Architecture of the MLP used as physics-informed neural network.}
	\label{fig:NN}
\end{figure}

To assimilate flow fields, such as velocity, temperature, or pressure fields, which are missing since they are not captured by a measurement, a multi-layer perceptron (MLP) is used. It is shown in figure~\ref{fig:NN} and serves as a universal function approximator for the $u,v,w,T,p$ properties of the flow.
The MLP consists of five fully connected hidden layers, each with a constant width of $N_\mathrm{N}$ neurons. This block of hidden layers is connected with linear scaling layers to the actual input and output to improve the condition of the network.

Each fully connected layer $\mathbf{g}^{(L+1)}$ is thereby configured as a linear combination, multiplying the weight matrix $\mathbf{W}$ with the output $\mathbf{g}$ of the previous layer $L$ and adding the bias vector $\mathbf{b}$, which is then activated by a non-linear function $\sigma$:
 \begin{equation}
	\mathbf{g}^{(L+1)}=\sigma(\mathbf{W}^{(L+1)}\cdot\mathbf{g}^{(L)}+\mathbf{b}^{(L+1)}).\label{eq:single_layer}
\end{equation}

Overall this network can be viewed as a function
 \begin{equation}
	 [u_x, u_y, u_z, T, p] = \mathbf{f}([t,x,y,z],\boldsymbol{\theta}) \label{eq:NN_fun}
\end{equation}
which is approximating solutions to the partial differential equations describing the investigated flow. Thereby, $\boldsymbol{\theta}$ denotes the complete set of the trainable parameters of the network, i.e. the weights and biases of the layers.

Regarding the activation function, the hidden layer neurons are uniformly activated with either a sine as a periodic function, a hyperbolic tangent, or an exponential linear unit (ELU) (see \cite{Clevert2015}) for comparison.
However, the hyperbolic tangent function is used as the activation for the scaled output layer in any case due to its asymptotic behavior for large absolute inputs.

\paragraph{Scaling layers}
Both scaling layers aim to condition the neural network in a way that its layer outputs are all scaled between $-1$ and $1$. Therefore, the already dimensionless input coordinates are scaled by

 \begin{equation}
	\zeta_\mathrm{scaled} = \frac{2}{\zeta_\mathrm{max}-\zeta_\mathrm{min}} \zeta + \frac{\zeta_\mathrm{min}+\zeta_\mathrm{max}}{\zeta_\mathrm{min}-\zeta_\mathrm{max}}, \label{eq:input scaling}
\end{equation}

with $\zeta \in \{t,x,y,z\}$ and $\zeta_\mathrm{min}$, $\zeta_\mathrm{max}$ being the boundaries of the examined domain.
This is especially necessary for the time coordinate, since we are examining relatively short intervals in the context of large DNS time coordinates \footnote{The scaling can also be performed as a data preprocessing step. Here, we opted to implement it as part of the PINN architecture for compatibility reasons.}.

For the output, only the temperatures are scaled with 
 \begin{equation}
	T = T_\mathrm{scaled}/2, \label{eq:output scaling}
\end{equation}
as $-0.5$ and $0.5$ are the dimensionless temperatures of the bottom and top surface, which limit the range for the fluid.
Since the velocity and pressure values do not have such clear limits, but are well distributed within $[-1,1]$, no further scaling is applied to them.
 
\subsection{Sampling}\label{sec:sampling}
\label{sec:ML_data}
The sampling for the training process is conducted according to the DNS grid. To limit the computational effort, only $11$ snapshots covering a time of $0.5\,t_\mathrm{ff}$ are processed.
Together with the full DNS grid resolution of $64^3$ used for the spatial sampling, this yields a number of approximately $2.9 \times 10^6$ data points.
Thus, for each epoch, randomly composed mini-batches of size $N_\mathrm{b} = 4\times10^3$ are provided for the individual training steps to benefit from the advantages of large batch sizes (\cite{Sankaran2022}).

\paragraph{Jittery collocation}
The collocation points of each training step, which are used to evaluate the PDEs, are based on the provided data points, but modified by some jitter $\delta_\mathrm{c}$. That means, in every training step, the dislocation $\delta_\mathrm{c}$ obtained from a random uniform distribution within $[-\Delta_\zeta/2,\Delta_\zeta/2]$ is added to the coordinates of the collocation points, where $\Delta_\zeta$ is the distance between two grid points for the coordinate $\zeta$.
This ensures that the PDEs are evaluated at a variety of positions between the original grid points over the course of the training, which is intended to prevent the formation of spurious subgrid-scale structures.
 
\paragraph{Boundary treatment}
In this assimilation approach, the boundary conditions are already partially covered by the provided data, which is also sampled at the boundaries.
To further support the assimilation, a separate boundary loss that covers the Dirichlet boundaries of the target fields has been implemented, i.e. the six walls for velocity assimilation or the constant temperature top and bottom plates for temperature assimilation.

These boundaries are sampled with a separate set of points. To account for the reduced density of data and collocation points used in the mini-batch of a training step, each boundary surface is represented by $8\times8$ spatial grid points for the same time instances provided by the data snapshots. Their spatio-temporal coordinates are then also modified by a jitter, as described above, to provide a dense boundary sampling by considering a large number of training steps.
In this way, the network is provided with $\tilde{\mathbf{u}}=0$ on all six faces of the cubic domain in the case of velocity assimilation and $\tilde{T}=\pm0.5$ for the bottom and top face in the case of temperature assimilation.

\subsection{Optimization}\label{sec:opti}
\label{sec:ML_opt}
To perform the training, the PINN is implemented in the framework of Tensorflow (\cite{Abadi2016}) and Keras (\cite{Chollet2015}).
More specifically, the Adam optimizer (\cite{Kingma2014}) is used in the default configuration, except for a constant learning rate of $10^{-4}$ for a fixed number of $5000$ epochs.
The overall loss function $\mathcal{L}$ used to train the network is a sum of several loss contributions described in more detail below.
First, the data loss $\mathcal{L}_\mathrm{data}$ is a mean-squared error loss of the provided data. That means, it is defined as $\mathcal{L}_\mathrm{data}=\mathcal{L}_T$ in the case of velocity assimilation and $\mathcal{L}_\mathrm{data}=\mathcal{L}_\mathbf{u}$ in the case of temperature assimilation:

\begin{eqnarray}
	\mathcal{L}_T &=& \frac{1}{N_\mathrm{b}} \sum_{j=1}^{N_\mathrm{b}} |T_j-\tilde{T}_j|^2 \label{eq:T_loss} \\
	\mathcal{L}_\mathbf{u} &=& \frac{1}{3 N_\mathrm{b}} \sum_{j=1}^{N_\mathrm{b}} \|(\mathbf{u}_j-\tilde{\mathbf{u}}_j)^2\|_1 \label{eq:u_loss} 
\end{eqnarray}

Regarding the physics losses, the contributions of the residuals of the Navier-Stokes equations ($\mathcal{L}_\mathrm{NSe}$), the convection-diffusion equation of the temperature ($\mathcal{L}_\mathrm{CDe}$) and the continuity equation ($\mathcal{L}_\mathrm{Coe}$) are calculated as shown below.

\begin{eqnarray}
	\mathcal{L}_\mathrm{NSe} &=& \frac{1}{3N_\mathrm{b}} \sum_{j=1}^{N_\mathrm{b}} \left\| \left( \frac{\partial \mathbf{u}_j}{\partial t} + (\mathbf{u}_j\cdot\nabla)\mathbf{u}_j + \nabla p_j -\sqrt{\frac{\mathrm{Pr}}{\mathrm{Ra}}}(\nabla\cdot\nabla)\mathbf{u}_j - \mathbf{e}_z T_j \right)^2\right\| _1 \label{eq:NS_loss} \\
	\mathcal{L}_\mathrm{CDe} &=& \frac{1}{N_\mathrm{b}} \sum_{j=1}^{N_\mathrm{b}}  \left|  \frac{\partial T_j}{\partial t} + (\mathbf{u}_j\cdot\nabla)T_j -\sqrt{\frac{1}{\mathrm{Pr}\mathrm{Ra}}}(\nabla\cdot\nabla)T_j\right|^2\ \label{eq:CD_loss}\\
	\mathcal{L}_\mathrm{Coe} &=& \frac{1}{N_\mathrm{b}} \sum_{j=1}^{N_\mathrm{b}}  \left| \nabla \cdot \mathbf{u}_j \right|^2\ \label{eq:Co_loss}  
\end{eqnarray}

Finally, the boundary losses $\mathcal{L}_\mathrm{bounds}$ are also data-based mean-squared error losses, and therefore, constructed as the data loss but for the respective other field and dedicated sampling points (cf. section~\ref{sec:sampling}).

The computation of the physics loss terms as well as the gradients of the losses with respect to the trainable parameters is facilitated by automatic differentiation (\cite{Baydin2018}).

\paragraph{Pressure centering}
An intricacy of finding solutions to the Navier-Stokes equations is the treatment of the pressure. Since the Boussinesq-approximated Navier-Stokes equations depend only on the pressure gradients, the absolute pressure level is not defined. This can lead to overall pressure levels close to $-1$ or $1$, which should be avoided due to the vanishing gradients caused by the activation of the scaled output layer, which is a hyperbolic tangent regardless of the case.
Therefore, we introduce another loss term $\mathcal{L}_\mathrm{pc}$, which centers the pressure values to variations around a level of $0$:

\begin{equation}
	\mathcal{L}_\mathrm{pc} = \left| \frac{1}{N_\mathrm{b}} \sum_{j=1}^{N_\mathrm{b}} p_j \right|  \label{eq:pc_loss}
\end{equation}

\paragraph{Loss component weighting}
Having different loss components means that the described training problem is based on a weighted-objective optimization. Therefore, weighting factors $\lambda_i$ are introduced for each loss term $\mathcal{L}_i$ to control their importance with respect to our reference loss contribution $\mathcal{L}_\mathrm{data}$, i.e. $\lambda_\mathrm{data} \equiv 1$.
Since the physical scales contribute to the weighting of the different loss terms, we consider the use of non-dimensional equations as a first step in weighting the different contributions.
Due to e.g. large gradients in the boundary layers, this per se does not guarantee PDE loss terms of an equal order of magnitude as the data losses. Yet, the order of magnitude of the bulk gradients is close to the dimensionless velocity, temperature, and pressure values for the moderate $\mathrm{Ra}$ and $\mathrm{Pr}$ numbers considered here. This implies that the meaning of the weights $\lambda_i$ is close to one of importance in the present case, while for different dimensionless numbers, it may also include the compensation of the gradients within the flow.

Since adaptive weighting schemes are still under extensive investigation (\cite{Wang2021,Wang2023}), this study of comparing different activations was conducted with a constant hierarchical weighting. 
Therefore, the $\lambda_i$ are set in relation to $\mathcal{L}_\mathrm{data}$, which was given the highest importance since the available data represent a direct access to the ground truth. This is especially the case for the DNS data set used here. For a possible application with data subject to uncertainty, this evaluation must certainly be lowered.

Second in the hierarchy is the loss of the PDE, which provides the essential information for the particular assimilation case. For the velocity assimilation case, the transport of the scalar $T$ carries information about the velocities. Therefore, the convection-diffusion equation of the temperature is used to infer $\boldsymbol{u}$ (cf. velocity measurements based on optical flow \cite{Horn1981}). 
In the case of temperature assimilation, the information harnessed to infer the temperature is mainly in the buoyancy-driven acceleration of the fluid. Hence, the force balance of the Navier-Stokes equations is more important in this case.
This means that $\lambda_\mathrm{CDe}=10^{-1}$ is assigned for velocity assimilation and $\lambda_\mathrm{NSe}=10^{-1}$ for temperature assimilation.

The next step in this hierarchy is occupied by the respective remaining loss contribution of those representing the Navier-Stokes and convection-diffusion equations. This means that $\lambda_\mathrm{NSe}=10^{-2}$ is assigned for the velocity assimilation and $\lambda_\mathrm{CDe}=10^{-2}$ for the temperature assimilation. Hyperparameter studies supporting this task-dependent relation between $\lambda_\mathrm{NSe}$ and $\lambda_\mathrm{CDe}$ can be found in appendix~\ref{sec:app_lambda}.

The lowest category is the same for both assimilation cases. It includes the loss contribution of the continuity equation ($\mathcal{L}_\mathrm{Coe}$), as higher weights for this loss tend to smooth the velocity fields (see also \cite{Lucor2022}). The boundary losses ($\mathcal{L}_\mathrm{bounds}$) are also weighted in this lowest category to compensate for the fact that they are sampled separately in each training step.
Finally, the pressure centering loss ($\mathcal{L}_\mathrm{pc}$) is also set to the lowest weight, which is sufficient to keep the pressure at a reference level.
Thus, the respective weights are set as follows $\lambda_\mathrm{Coe}=\lambda_\mathrm{bounds}=\lambda_\mathrm{pc}=10^{-3}$.

\paragraph{Initialization}

When using periodic activation functions, the initialization of the weights is of paramount importance for a successful training result (\cite{Sitzmann2020}), since they determine the spatial and temporal wave numbers of the output of individual neurons. Therefore, we use the initialization of random weights from a uniform distribution in the interval $[ -\sqrt{{6}/{N^{(L-1)}}},\sqrt{{6}/{N^{(L-1)}}}] $ proposed by \cite{Sitzmann2020}, where $N^{(L-1)}$ is the number of neurons of the preceding layer.

As also discussed by \cite{Sitzmann2020}, the first hidden layer appears suitable to introduce higher spatial and/or temporal frequencies into the neural network.
Thus, we introduce these higher frequencies or wave numbers by initializing the weights of the first layer with random numbers from an interval $[ -w\sqrt{{6}/{N^{(0)}}},w\sqrt{{6}/{N^{(0)}}}] $. Here, $w$ serves as a factor to broaden the distribution of the initial weights and frequencies.
Unless otherwise specified, $w=4$ was used when sine activation functions were used.
A parameter study on the influence of the factor $w$ is also part of the appendix~\ref{sec:app_ini}.

 \subsection{Evaluation metrics}\label{sec:metrics}

We use the DNS dataset as ground truth to monitor the evolution of the mean average error $\mathrm{MAE}_\xi$ and Pearson's correlation coefficient $\mathrm{PCC}_\xi$ with $\xi\in \{u_x,u_y,u_z,T\}$. Both are defined as follows:

\begin{eqnarray}
	\mathrm{MAE}_\xi &=& \frac{1}{N_\mathrm{m}} \sum_{k=1}^{N_\mathrm{m}} |\xi_k-\tilde{\xi}_k| \label{eq:mae} \\
	\mathrm{PCC}_\xi &=& \frac{N_\mathrm{m} \sum_{k=1}^{N_\mathrm{m}} \xi_k\tilde{\xi}_k - \sum_{k=1}^{N_\mathrm{m}} \xi_k \sum_{k=1}^{N_\mathrm{m}} \tilde{\xi}_k }{\sqrt{N_\mathrm{m} \sum_{k=1}^{N_\mathrm{m}} \xi_k^2 - \left( \sum_{k=1}^{N_\mathrm{m}} \xi_k \right)^2 }\sqrt{N_\mathrm{m} \sum_{k=1}^{N_\mathrm{m}} \tilde{\xi}_k^2 - \left( \sum_{k=1}^{N_\mathrm{m}} \tilde{\xi}_k \right)^2 }}  \label{eq:pcc} 
\end{eqnarray}

Here, $N_\mathrm{m}$ is the number of samples used for the calculation. For this purpose, every fourth grid point of each spatial direction was sampled for each snapshot.

\section{Assimilation results}
\label{sec:results}

Since there are more studies on velocity assimilation in the literature, we will first discuss how the here presented method performs for the velocity assimilation in section \ref{sec:vel_assi}.
After that, the inverse task of temperature assimilation and its characteristics are discussed in section \ref{sec:T_assi}.

\subsection{Velocity assimilation}\label{sec:vel_assi}
\paragraph{Training process}

Focusing on the influence of the applied activation functions, different numbers of neurons of each layer and the various activations were tested.
For each combination, 3 training runs with different initialization were performed.

\begin{figure}[!h]
	\centering
	\includegraphics[width=0.99\linewidth]{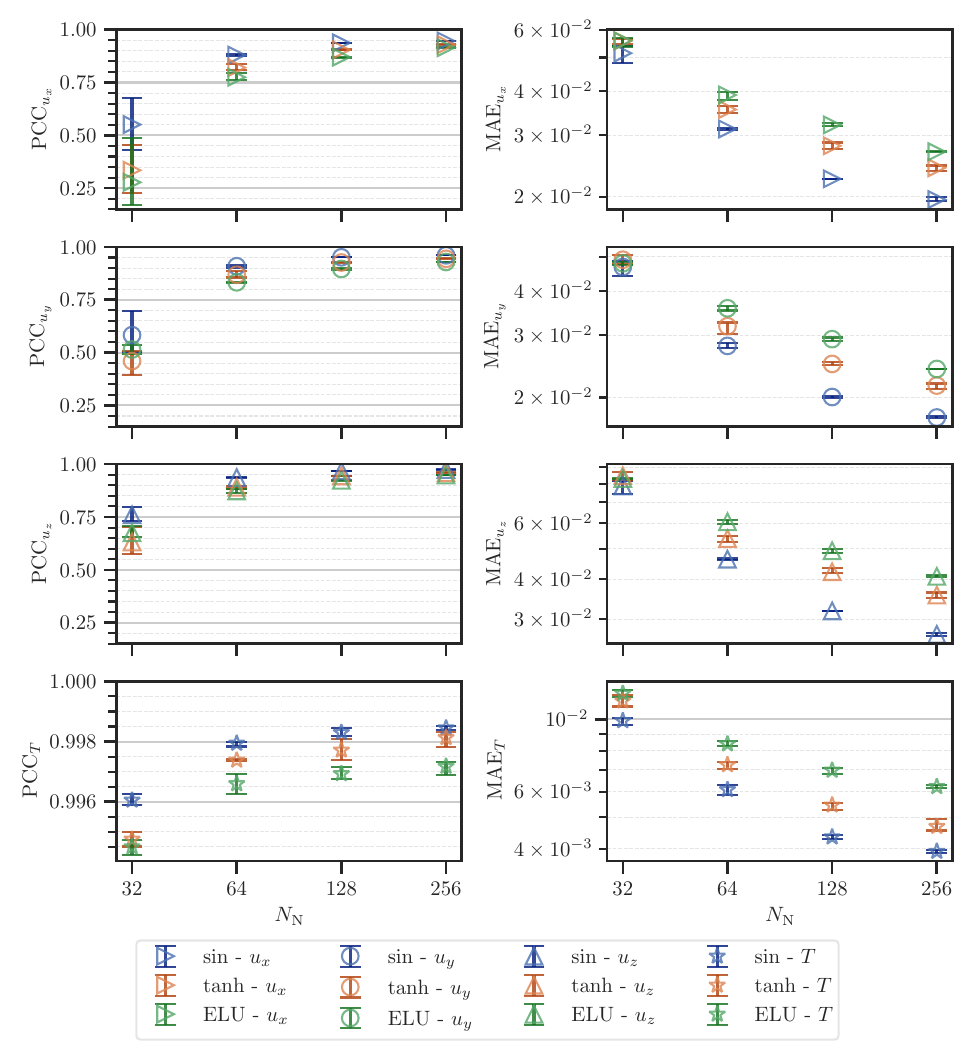}
	\caption{Correlation (left) and MAE (right) metrics of the velocity and temperature fields (top to bottom) achieved after 5000 epochs of velocity-assimilation training with varying activation functions and sizes ($N_\mathrm{N}$) of the hidden layers. The markers represent averages over multiple training runs and the error bars the respective minima and maxima.}
	\label{fig:uassi_size}
\end{figure}

Fig.~\ref{fig:uassi_size} shows the values of the validation metrics described in section~\ref{sec:metrics} obtained for the different training runs at the end of the 5000th training epoch.
While the correlation coefficients ($\mathrm{PCC}$) indicate how well the PINNs are able to model the existing flow structures, the additionally shown mean average errors ($\mathrm{MAE}$) also reflect the reproduction of the correct amplitudes.
The symbols represent the mean values of the three training runs, while the error bars display the respective minimum and maximum values.

Considering the correlation metric on the left side of figure~\ref{fig:uassi_size},
it is found that for all activation functions and network sizes $N_\mathrm{N}$ an almost perfect correlation of $\mathrm{PCC}_T > 0.99$ is obtained for the fit of the provided temperature fields.
On the other hand, the correlation of the assimilated velocity fields depends on $N_\mathrm{N}$ because there is an increase of the correlation with the number of neurons of the network for all activation functions.

Another trend shown by correlation values of the velocities in the different cases is that the correlation coefficients for the different components are always sorted from $u_z$ with the highest values to $u_x$ with the lowest values. This indicates the existence of characteristics inherent to the studied flow that make it easier to predict individual components, since the differences also occur in the two horizontal components.

To compare periodic and non-periodic activation functions, the correlation results for $u_x$ with sine and hyperbolic tangent activations are also highlighted in the inserted plot.
It shows that a periodic activation performs better than a non-periodic one for all network sizes.
In particular, sine-activated networks with a layer width of $128$ are reliable able to achieve correlations above $0.9$ for all velocity components.
It is also noteworthy that a sine-activated network with a layer width of $64$ performs similarly to a hyperbolic tangent network with twice the number of neurons.

The results for the MAE metric displayed on the right side of fig.~\ref{fig:uassi_size} confirm the above findings with lower errors for an increased number of neurons and the same hierarchy of activation functions.
In the best case, the MAE amounts to between $0.018$ and $0.027$ for the different velocity components.
However, it is noteworthy that the fit error of the provided temperature field also depends on the varied parameters, although the correlation metric did not show large differences. 

Also, doubling the number of neurons to $256$ yields only marginal benefits in terms of the correlation performance. Based on this, the case of $N_N=128$ will be investigated in more detail below as it represents the parsimonious case for this flow setup.  

\begin{figure}[!ht]
	\centering
	\includegraphics[width=0.99\linewidth]{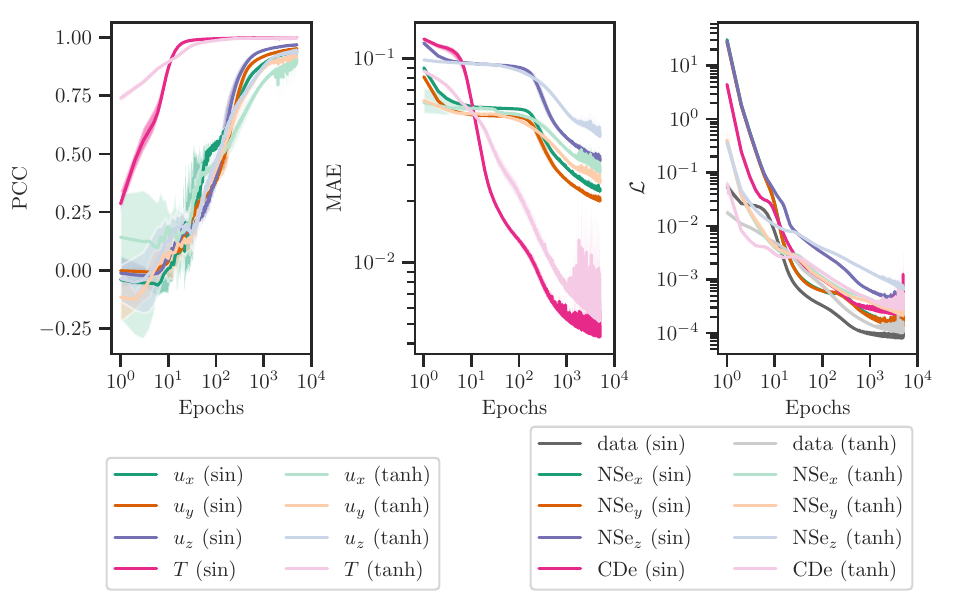}
	\caption{Evolution of the correlation (left), MAE (center) and losses (left) during the training. Darker lines represent the sine-activated cases and lighter lines hyperbolic tangent-activated ones. Each solid line describes the mean of the single velocity-assimilation training runs at $N_\mathrm{N}=128$, while the respective transparent envelope is bound by the minimum and maximum values.}
	\label{fig:uassi_train}
\end{figure}

To examine these results in more detail, figure~\ref{fig:uassi_train} shows the convergence plots of the discussed metrics (left and center) and a selection of loss terms (right) over the course of training.
The lines represent the mean values obtained in three different runs. The respective envelopes of the minima and maxima are represented by the shaded areas.
Thus, we can compare the periodic sine activation (darker colors) with the hyperbolic tangent activation (lighter colors), which represents the class of non-periodic activation functions, both for $N_\mathrm{N}=128$.
Due to the random initialization, there are significant differences in the correlations metric of the velocities at the beginning of the training.
However, these differences disappear during the first 100 epochs. 
Subsequently, they remain negligibly small for both activation functions.

Overall, the training process can be divided into several periods. During the first period, which lasts until about the 20th epoch, the training mainly optimizes the fit of the MLP to the provided  temperature fields. In this phase, the physics losses are reduced due to a general attenuation of the initial gradients in the inferred fields. 
After that, the temperature fields are well enough represented by the MLP, so that the reduction of the physics losses starts to improve the inferred velocities.
After a few hundred epochs, the correlation coefficient for the inferred velocity components starts to converge, meaning that the PINN has recovered the most salient velocity structures. This also marks the beginning of the decrease of the associated MAEs, showing that the physics loss terms are not only able to reconstruct the structures, but also push the inferred fields towards the quantitatively correct ones.
From about epoch 1000 onward the MAEs and losses appear to flatten out, while they also begin to oscillate.

In summary, both sets of runs would yield negligible improvements in the validation metrics after the observed training period.
The difference between the periodic and non-periodic activation functions is that the sine activation is able to fit the provided temperature field faster (except for the first 10 epochs) and better.
We attribute this to the periodic nature of the sine, which provides more complexity for a given number of neurons.
This is beneficial for the inferred fields in two ways: First, because they also require a certain amount of complexity provided by the MLP to be reconstructed, and second, as a better fit to the given temperature fields yields more accurate gradients to the physics losses.

\paragraph{Inferred fields}
\begin{figure}[!h]
	\centering
	\includegraphics[width=0.99\linewidth]{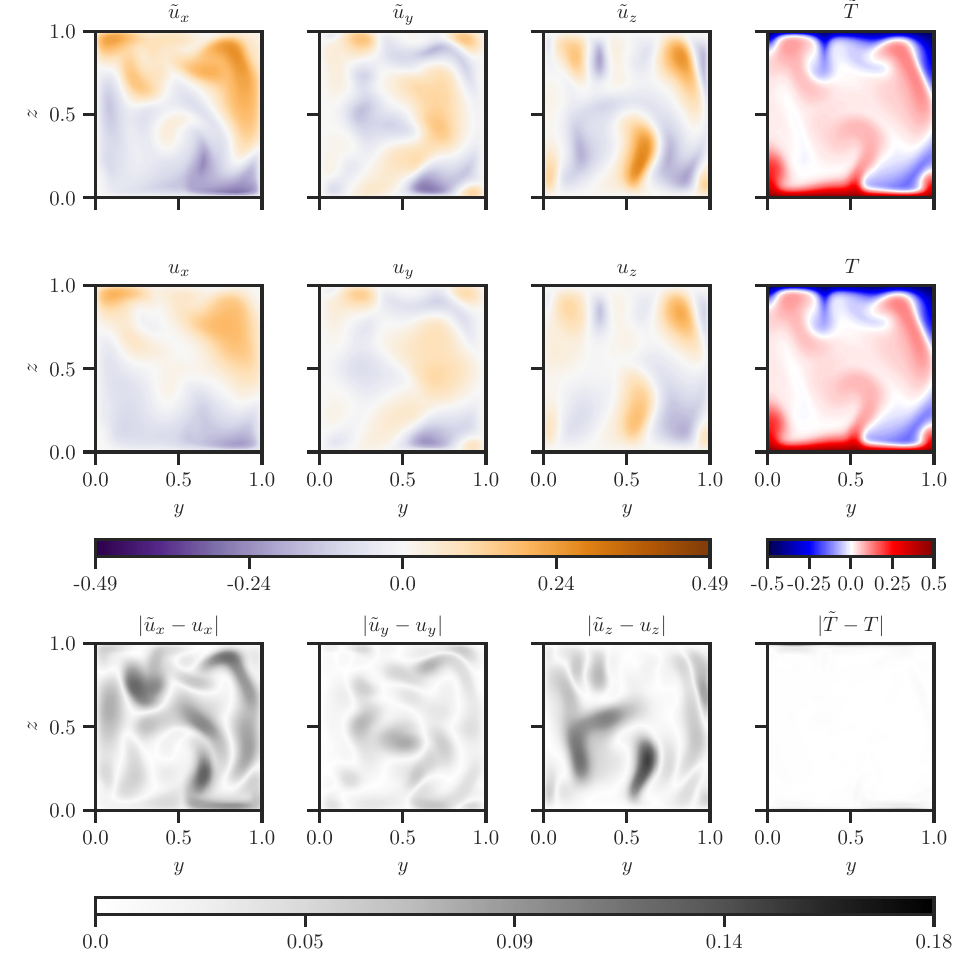}
	\caption{Comparison of the ground truth (top row) and PINN-generated (middle row) fields of the velocity components $u_x$, $u_y$, $u_z$ and the temperature $T$ in a central vertical cross-section for velocity assimilation with sine activation and $N_\mathrm{N}=128$.
	The bottom row comprises the contour plots of the respective absolute differences.}
	\label{fig:uassi_fields}
\end{figure}

Fig~\ref{fig:uassi_fields} shows a visual comparison of the ground truth fields (top row) with the inferred fields (middle row) in a vertical cross-section of the domain at $x\approx0.5$, obtained with sine activation and $N_\mathrm{N}=128$.  Additionally, the absolute differences between the top and middle fields are shown in the bottom row.
While the differences (bottom row) between the temperature fields of the PINN and the ground truth are close to zero, the velocity differences are clearly visible.

However, looking at the structures in the top and middle rows, it is found that they are qualitatively similar, although the amplitudes are different.
This is confirmed by the plots of the absolute differences, where the largest deviations occur at the same positions as the high velocity magnitudes in the ground truth data set.
To further investigate these differences in velocities between the PINN reconstruction and the ground truth, we subsequently analyze how well the PINN is able to reproduce extreme values and small structures.

\begin{figure}[!ht]
	\centering
	\includegraphics[width=0.99\linewidth]{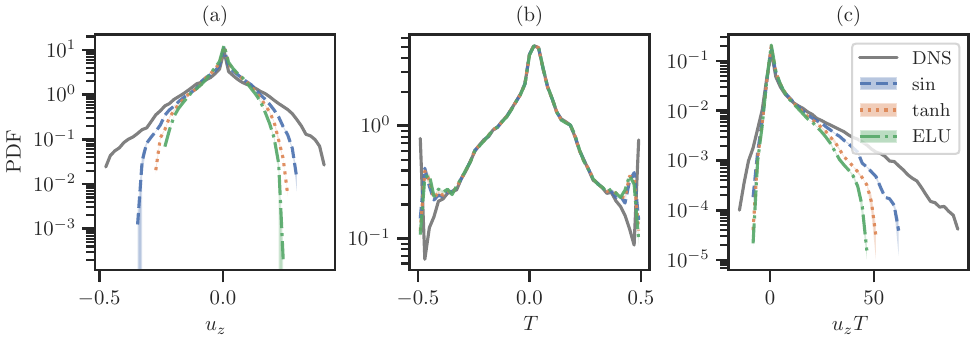}
	\caption{Global histograms of the vertical velocity component $u_z$ (a), the temperature $T$ (b) and the vertical convective heat flux $u_z T$ (c) for the velocity assimilation cases with $N_\mathrm{N}=128$. The lines represent an average of the respective training runs, while the envelopes are bound by the minimum and maximum values.}
	\label{fig:uassi_histo}
\end{figure}

The reproduction of extreme values is considered by analyzing the global histograms of vertical velocity ($u_z$), temperature ($T$) and vertical convective heat transport ($u_z T$) displayed in fig.~\ref{fig:uassi_histo}.
While the gray lines represent the ground truth data, the colored lines correspond to the differently activated PINN models for $N_\mathrm{N}=128$.

First, we consider the temperature data displayed in fig.~\ref{fig:uassi_histo}\,(b) as it concerns the data provided to train the PINN.
Besides a central peak, the distribution of the ground truth temperature also shows additional peaks at $T=\pm0.5$, which refer to the constant temperature boundary conditions.
This distribution is closely reproduced by the PINN models regardless of the activation function, except for the regions near $T=\pm0.5$.
There, all cases show a shift of the extreme boundary values towards zero, showing that the extreme gradients within the boundary layer are difficult to model with a PINN, even if the corresponding data are available.

Regrading $u_z$ (fig.~\ref{fig:uassi_histo}\,(a)), the histograms of all PINN models are significantly narrowed, i.e. they underestimate the occurrence of extreme values. This is consistent with the lack of amplitude visible in the center sections presented in fig.~\ref{fig:uassi_fields}.
However, this effect is less pronounced for the sine activation than for the non-periodic functions, demonstrating that the PINN's ability to model the amplitudes of the velocity vector field benefits from a periodic activation.

In addition, the PDFs of the convective heat fluxes $u_zT$ are displayed in  figure~\ref{fig:uassi_histo}\,(c).
There, the distributions of the PINN-modeled heat fluxes display the same narrowed characteristic as for the vertical velocity component.
However, the main differences between the various activation functions occur for the positive tail of the distribution, which constitutes the strong positive skew of the ground truth.
Again, In this regard, the sine activated PINNs perform best again, as they model the right tail the closest.

\begin{figure}[!ht]
	\centering
	\includegraphics[width=0.99\linewidth]{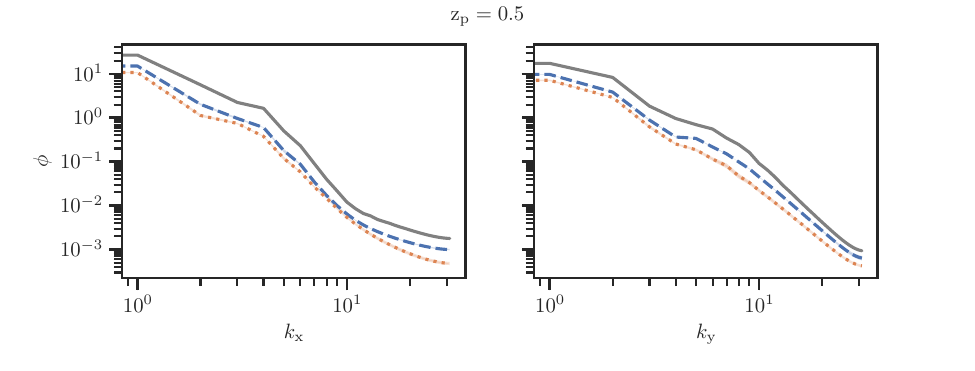}	\includegraphics[width=0.99\linewidth]{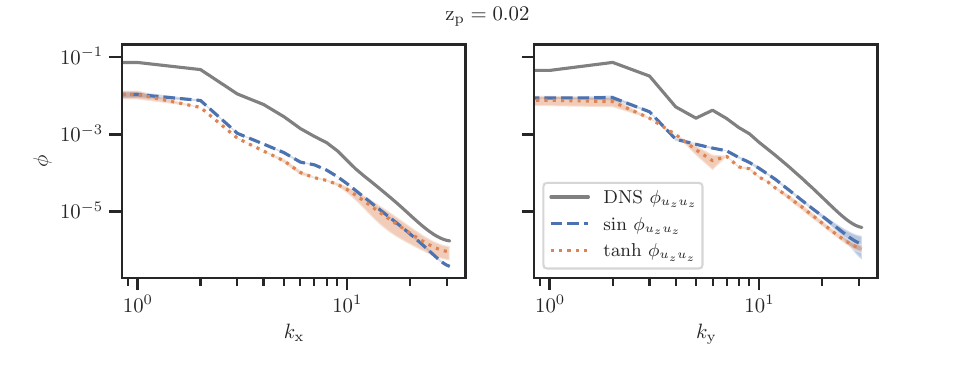}
	\caption{Comparison of the spectra in $x$- and $y$-directions of the $u_z$ field of the ground truth (gray) and sine (blue) and hyperbolic tangent (orange) activation with $N_\mathrm{N}=128$. The top row comprises the spectra in the central horizontal plane and the bottom row for a plane close to the boundary, while the columns comprise the different directions $k_x$ (left) and $k_y$ (right). The lines display the average over the runs while the envelopes are bound by the respective minima and maxima.}
	\label{fig:uassi_fft}
\end{figure}

To test the ability to reproduce the turbulence structure in the assimilated velocity field, power density spectra $\phi_{u_z u_z}(k_\zeta)$ are shown in fig.~\ref{fig:uassi_fft} in a central horizontal plane $z=0.5$ (top) and a parallel plane near the heated bottom plate $z=0.02$ (bottom).
The spectra also distinguish between the wave numbers $k_x$ (left) and $k_y$ (right) in the two planar directions.
For each of these positions, the figure displays the spatial spectra evaluated with the DNS data and the inferred fields using sine and hyperbolic tangent activations at $N_\mathrm{N}=128$.
Note that these results are obtained by applying Fast Fourier Transforms (FFTs) to non-periodic signals, which means that they are affected by spectral leakage.
This is particularly noticeable for some spectra that flatten out at high wave numbers. This effect is mitigated by the no-slip boundary conditions, which impose something like a natural window function.
However, since this leakage is present in both the ground truth and inferred field data, the comparison is still possible.

In this comparison, the spectra of the PINN-generated fields exhibit lower power densities overall, regardless of plane, direction, and wave number.
Of the two activation functions shown, the sine activation generally exhibits smaller differences to the ground truth of the spectral power densities than the hyperbolic tangent activation, confirming the observations made with respect to figures~\ref{fig:uassi_fields} and \ref{fig:uassi_histo}.

However, it cannot be observed that the gap of the power density spectra between the PINN models and the ground truth would grow towards higher wave numbers $k$. This means that the impression of the inferred velocity fields in fig.~\ref{fig:uassi_fields} is not to be understood as a blurring, as this would result in small scales being suppressed more than the larger scales. On the contrary, the softer appearance of the inferred fields is caused by a general suppression of the amplitudes.
For the plane at $z\approx0.02$ , the gap between ground truth and inferred vertical velocity actually closes for large wave numbers. This may be caused by an underrepresentation of the large scales in the inferred vertical velocity $u_z$, while the existing smaller structures are still well represented.

With the PINN, which minimizes the losses in form of the residuum of the Navier-Stokes equations, the pressure fields are also obtained during the assimilation. 
We have chosen to omit their investigation in the main part of this paper in order to keep the focus on temperature and velocity, which contribute to the convective heat transport. 
For insights into the inferred pressure fields, we refer the reader to appendix~\ref{sec:app_press}.

\subsection{Temperature assimilation}\label{sec:T_assi}
\begin{figure}[!ht]
	\centering
	\includegraphics[width=0.99\linewidth]{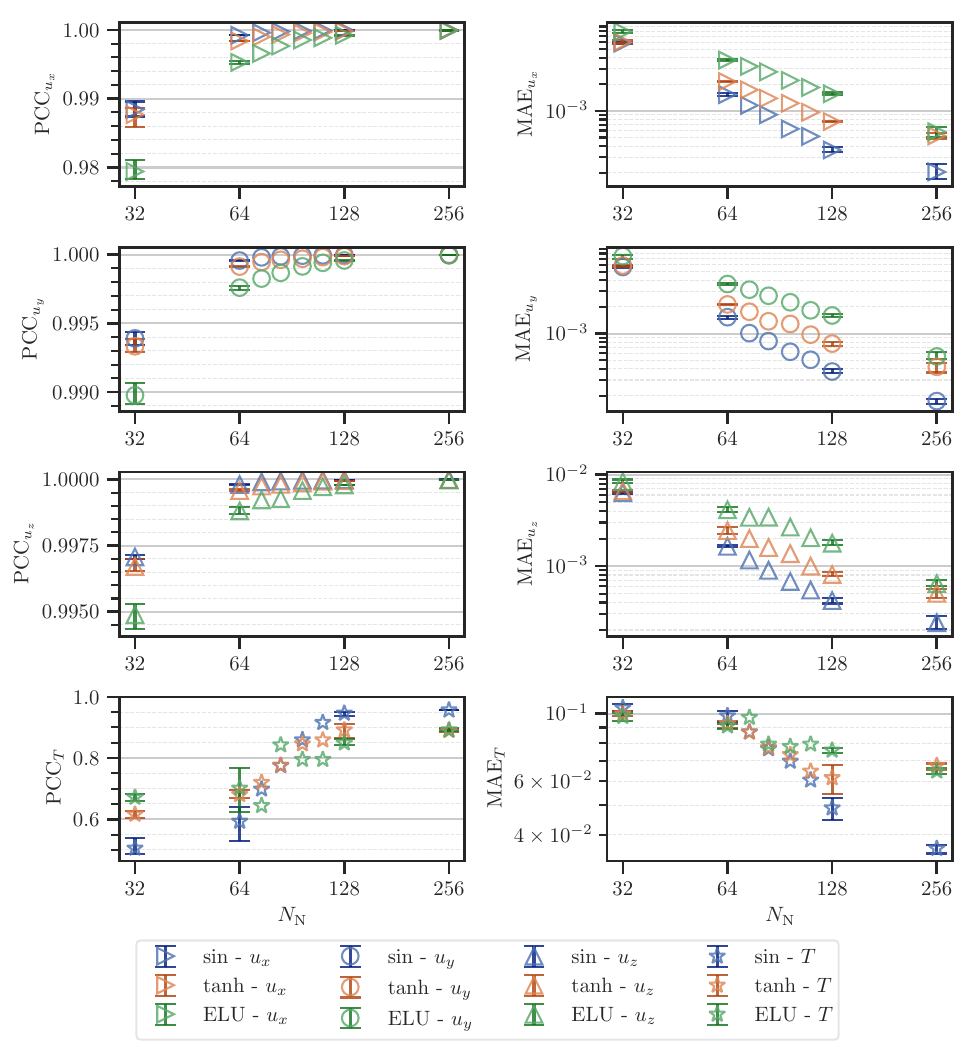}
	\caption{Correlation (left) and MAE (right) metrics of the velocity and temperature fields (top to bottom) achieved after 5000 epochs of temperature-assimilation training with varying activation functions and sizes ($N_\mathrm{N}$) of the hidden layers. The markers represent averages over multiple training runs and the error bars the respective minima and maxima.}
	\label{fig:Tassi_size1}
\end{figure}

\paragraph{Training process} 
Following the structure of the discussion of the above section, which deals with velocity assimilation, an overview of the performance of the temperature assimilation with different activation functions and neuron numbers $N_\mathrm{N}$ is shown in fig.~\ref{fig:Tassi_size1}.
Here, all combinations of neuron number\footnote{We limit our variation of the neuron number to changes of the width of the hidden layers to keep the investigations concise. Another option would be a variation of the number of hidden layers. A brief overview of the effects of this hyperparameter can be found in appendix~\ref{sec:app_depth}.} and activation function were analyzed based on three training runs. The only exception is the best case, $N_\mathrm{N}=128$ with sine activation, where five training runs were performed. The displayed markers are again averages over the training runs, and the error bars represent the respective minima and maxima.

Overall, the two plots display the trend that the correlation coefficients (fig.~\ref{fig:Tassi_size1}, left side) increase and the absolute errors decrease (fig.~\ref{fig:Tassi_size1}, right side) with increasing neuron number as it was already observed for the velocity assimilation.
To investigate the characteristics of the temperature assimilation task, we first discuss the correlation coefficients in more detail.
Starting with $N_\mathrm{N}=32$, the achieved correlations are between $0.5$ and $0.7$, which is rather low. At this stage, the sine activated neural network has the lowest correlation coefficient when comparing the activation functions, which means that it performs the worst in this configuration, while ELU performs the best.

For networks twice the size ($N_\mathrm{N}=64$), the increase in correlation values is not as great as for the velocity assimilation. More strikingly, the variations between the different runs of both sine and ELU activations increase with this doubling of $N_\mathrm{N}$.

For the next doubling of $N_\mathrm{N}$ to $128$, there is a transition where the PINN models reach correlations of $\rho>0.8$ for all activation functions with sine now giving the best results with $\rho>0.94$. Also, the initialization-induced variations of the results decrease again.
To highlight this transition to viable inference results, the metrics for individual intermediate training runs for $N_\mathrm{N} \in \{74,84,97,111\}$ have been added to the plots, where they show the sine activation surpassing the other activation functions for the temperature metrics.
The next doubling to $N_\mathrm{N} = 256$ offers only small improvements, confirming that the relevant transition takes place between $64$ and $128$ neurons per hidden layer. As for the velocity assimilation, this renders $N_N=128$ as parsimonious for the given task, so the detailed analysis is performed for this neuron number.

Regarding the fit of the provided velocity fields, correlation coefficients close to $1$ are obtained in all cases.
However, the associated MAE values (fig.~\ref{fig:Tassi_size1}, right side) decrease with increasing neurons.
Regardless of the number of neurons, the lowest absolute errors are again obtained with sine activation, confirming similar observations for the velocity assimilation.

Compared to the qualitative differences of the velocity errors obtained for the different activation functions, the differences between activations for the assimilated temperature fields are smaller (in terms of logarithmic scaling), but at a higher level overall.
However, they follow the same trends as the correlation coefficients with the sine activation reaching $\mathrm{MAE}_T\approx0.049$ at $N_\mathrm{N}=128$ and $\mathrm{MAE}_T\approx0.036$ at $N_\mathrm{N}=256$.

\begin{figure}[!h]
	\centering
	\includegraphics[width=0.99\linewidth]{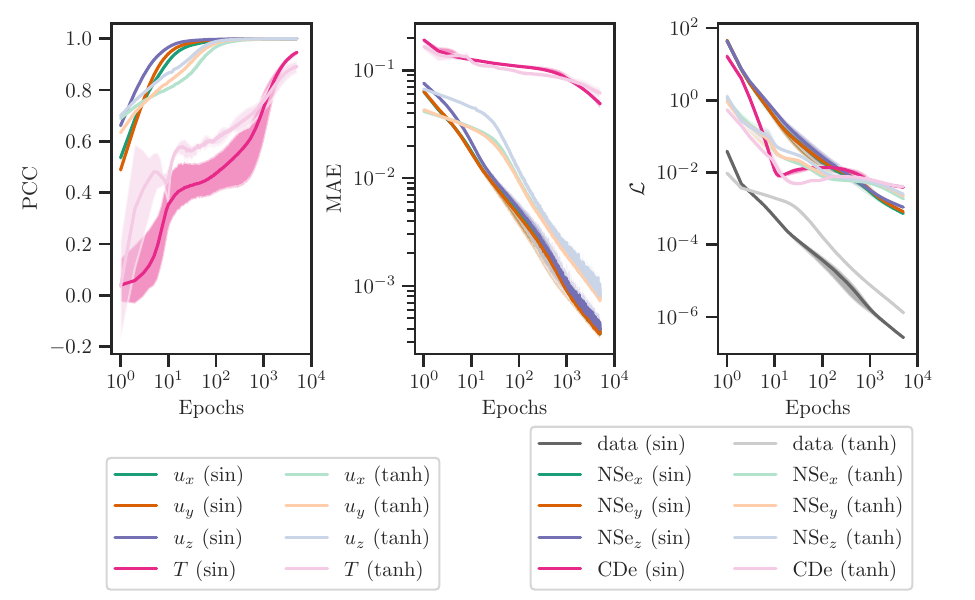}
	\caption{Evolution of the correlation (left), MAE (center) and losses (left) during the training. Darker lines represent the sine-activated cases and lighter lines hyperbolic tangent-activated ones. Each solid line describes the mean of the single temperature-assimilation training runs at $N_\mathrm{N}=128$, while the respective transparent envelope is bound by the minimum and maximum values.}
	\label{fig:Tassi_train}
\end{figure}

To investigate the performance of the training processes for the different activation functions, fig.~\ref{fig:Tassi_train} shows the correlation and MAE metrics along with selected losses for the training process of test cases with sine (darker) and hyperbolic tangent (lighter) activation at $N_\mathrm{N}=128$. The lines represent averaged values over the five or three test runs, respectively, while the shaded area represents the respective envelope between minima and maxima.

As a first observation, these envelopes cover only insignificantly small regions with the exception of the correlation metric (left side of fig~\ref{fig:Tassi_train}) of the inferred temperatures. There, variations of up to $0.2$ can occur between individual training runs. However, this envelope still collapses towards the end of the training at 5000 epochs, meaning that the final results do not depend on a favorable random initialization.

Regarding the correlation progress of the training with respect to the fields fitted to the ground truth data \textemdash\ here, the velocity fields \textemdash\ the sine activation produces high ($\mathrm{PCC}_u>0.9$) correlation values significantly faster than the hyperbolic tangent activation. This confirms the finding from the velocity assimilation, where sine-activated PINNs are also faster in fitting the provided data.  

Next, the evolution of the correlation coefficients for the assimilated temperature fields is discussed: They initially grow strongly to $\mathrm{PCC}_T\approx0.4$ for both activation functions considered. After that, the increase of the correlation value is slower, before the values grow faster again (on a logarithmic epoch scale) at about 1000 epochs.
This second growth period is significantly stronger for the sine activation, i.e. it surpasses the performance of the hyperbolic tangent during this period.

These two periods of faster growth separated by a slower one are also obtained for the MAE metrics (center plot of fig.~\ref{fig:Tassi_train}) of the assimilated temperatures.

Considering the velocity MAEs and the losses (right side of fig.~\ref{fig:Tassi_train}), decays are observed towards the end of the training, while for the velocity assimilation they already started to level off after about 1000 epochs.
This means that PINNs are harder to train for temperature assimilation than for velocity assimilation because the $5000$ epochs are not sufficient to achieve a leveling out as in the velocity assimilation case.
A fixed limit of $5000$ epochs is still justifiable, since the logarithmic epoch axis still means strongly diminishing returns for further training, especially regarding the correlation values.

\paragraph{Inferred fields}
\begin{figure}[!h]
	\centering
	\includegraphics[width=0.99\linewidth]{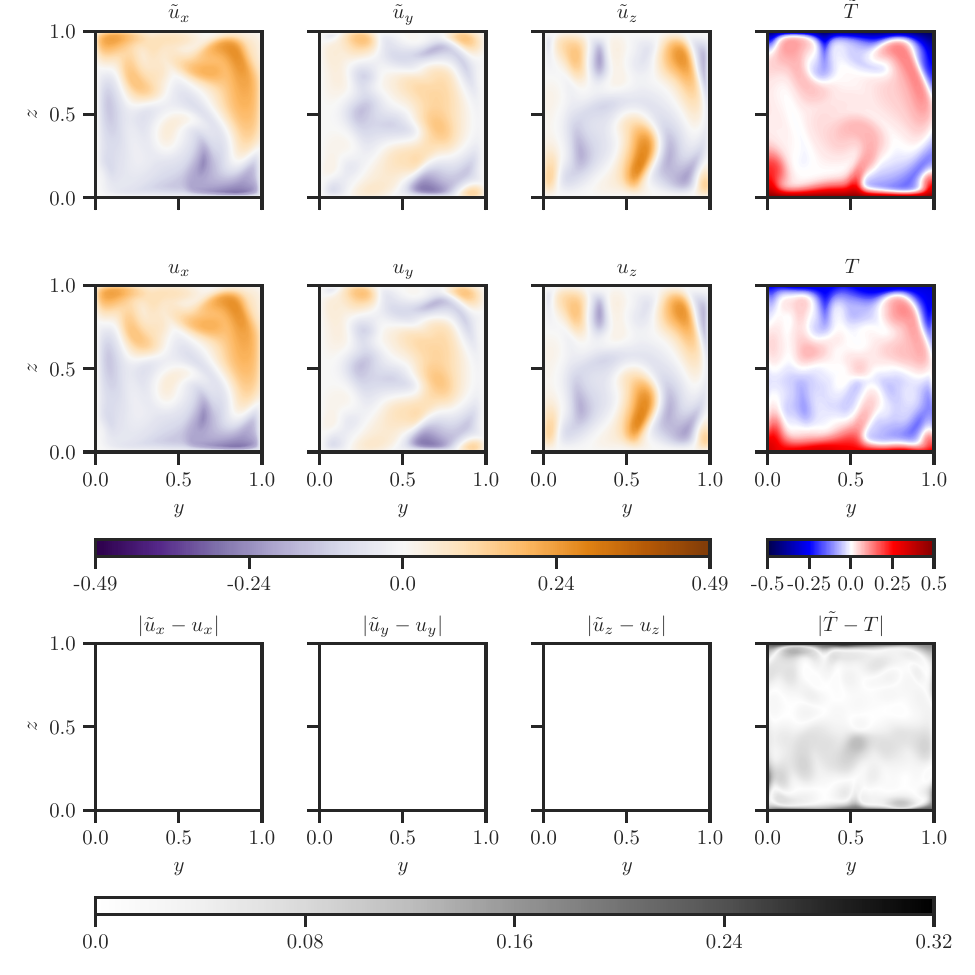}
	\caption{Comparison of the ground truth (top row) and PINN-generated (middle row) fields of the velocity components $u_x$, $u_y$, $u_z$ and the temperature $T$ in a central vertical cross-section for temperature assimilation with sine activation and $N_\mathrm{N}=128$.
	The bottom row comprises the contour plots of the respective absolute differences.}
	\label{fig:Tassi_fields}
\end{figure}

The comparison between exemplary fields generated by the sine-activated PINN at $N_\mathrm{N}=128$ and the ground truth is displayed in figure~\ref{fig:Tassi_fields}.
It includes velocity and temperature fields in a vertical central section for the ground truth (top row) and the PINN (middle row) as well as the resulting absolute differences (bottom row).

Similar to the case of velocity assimilation, the fit of the fields for which data were provided to train the PINN shows hardly any noticeable absolute errors. 
However, this is not the case for the inferred temperature field. 
While the PINN is able to reproduce all significantly warm or cold structures, such as the boundary layers and the plumes originating from them, it particularly struggles in the bulk region. This especially applies to the region of $0.25<z<0.5$ for the displayed vertical center section. Since the actual temperature amplitudes are smaller in this region, so is the resulting buoyancy acceleration. This finally makes an accurate reconstruction of the temperature more difficult.

\begin{figure}[!h]
	\centering
	\includegraphics[width=0.99\linewidth]{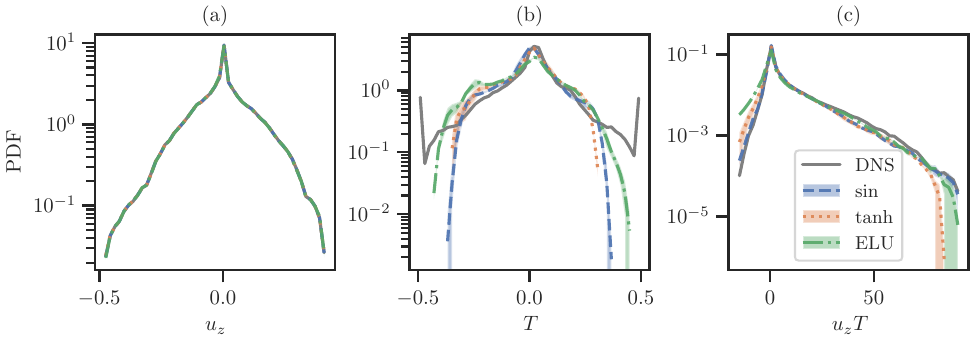}
	\caption{Global histograms of the vertical velocity component $u_z$ (a), the temperature $T$ (b) and the vertical convective heat flux $u_z T$ (c) for the temperature assimilation cases with $N_\mathrm{N}=128$. The lines represent an average of the respective training runs, while the envelopes are bound by the minimum and maximum values.}
	\label{fig:Tassi_histo}
\end{figure}

To investigate how the described differences between the ground truth and the inferred fields manifest themselves in the flow statistics, the global histograms of $u_z$, $T$ and $u_z T$ are compared for the different activation types at $N_\mathrm{N}=128$ in fig.~\ref{fig:Tassi_histo}.

As expected from the results discussed above, the differently activated PINNs all manage to reproduce the distribution of the vertical velocity, shown in figure~\ref{fig:Tassi_histo}\,(a), without significant differences.

Regarding the histograms of the inferred $T$ fields shown in figure~\ref{fig:Tassi_histo}\,(b), the differently activated PINNs also follow the expectations based on the velocity assimilation and are not able to cover the whole width of the temperature range, leading to slightly higher values of the probability density function for moderate temperature amplitudes.
While the sine and hyperbolic tangent activations perform almost equally, the ELU activation is able to cover the widest temperature range. Yet, it also underestimates the occurrence of temperatures close to $T=0$.

The histogram of $u_z T$ in fig.~\ref{fig:Tassi_histo}\,(c) shows the characteristic right-skewed distribution of convective heat fluxes for the ground truth and the investigated PINN models. 
While the ELU-activated PINN was able to produce the widest temperature distribution, it also exhibits the largest deviation of the tested activations for the negative tail of $u_z T$, where all activations overestimate the occurrence of locally negative convective heat fluxes. We attribute this to an inherent mismatch between the large temperature amplitudes and the corresponding vertical velocities.
In contrast, both tails of the histogram $u_z T$ are close to the ground truth for the sine activation.
The hyperbolic tangent activation shows similar deviations to the ELU activation, but with more underestimation of positive values and less overestimation of negative values.
This illustrates that the distributions of $u_z$ and $T$ alone are not sufficient to evaluate the quality of the predictions.

Overall, the inferred convective heat flux histogram is in much better agreement with the ground truth for the temperature assimilation than for the velocity assimilation.
This is due to the fact that the largest temperature deviations occur in regions where the convective heat flux is low, namely the bulk region with low temperature amplitudes and the boundary layers with low vertical velocities.

\begin{figure}[!h]
	\centering
	\includegraphics[width=0.99\linewidth]{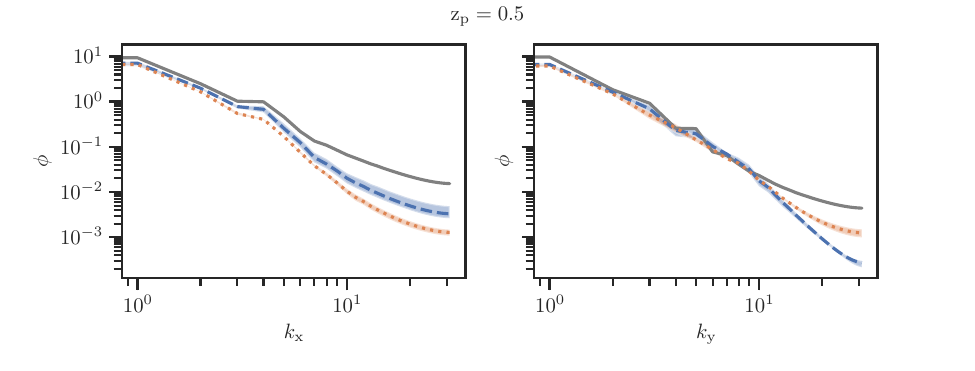}	\includegraphics[width=0.99\linewidth]{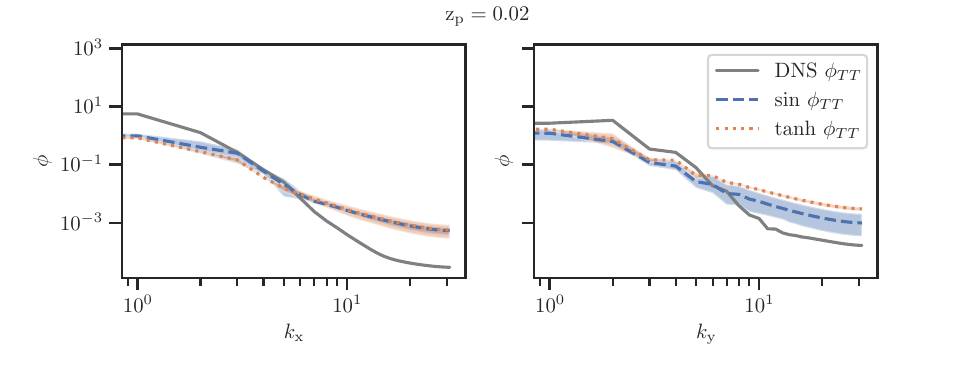}
	\caption{Comparison of the spatial spectra of the ground truth (gray) and the inferred temperature fields for sine (blue) and hyperbolic tangent (orange) activation. The top row comprises the spectra a the central horizontal plane and the bottom row for one close to the boundary, while the columns comprise the different directions $k_x$ (left) and $k_y$ (right). Data shown for the training runs with $N_\mathrm{N}=128$. The lines display the average over the runs while the envelopes are bound by the respective minima and maxima.}
	\label{fig:Tassi_fft}
\end{figure}

The power density spectra of the temperature fields are shown in fig.~\ref{fig:Tassi_fft} for the $x$ (left) and $y$ (right) directions in the planes $z=0.5$ (top) and $z=0.02$ (bottom), to investigate the ability of the differently activated PINNs to assimilate structures of different sizes.
Following the systematics of the comparable fig.~\ref{fig:uassi_fft}, the sine and hyperbolic tangent activations represent both periodic and non-periodic activations for the displayed cases of $N_\mathrm{N}=128$.

In the bulk flow at $z=0.5$, the inferred temperature fields of both activations, sine and hyperbolic tangent, show that the gap of spectral power density widens towards higher wave number, implying a deficiency in the reproduction of smaller structures.
Since none of the power densities predicted with the activation functions agrees well with the ground truth in either direction, there is no clear advantage for one of the two activations.
A possible explanation for this shortcoming could be numerical effects that add to the thermal diffusivity.
However, the power density spectra for the vicinity of the bottom plate ($z=0.02$) question this explanation, as they both exhibit an overestimation of power densities for large wave numbers.
This translates to an overestimation of the temperature variation on the small scales while the variations on the larger scales are still underestimated.
Therefore, we explain this by smaller structures leaking in from the region farther from the bottom plate, while large scale structures are still suppressed because the PINNs fall short on reproducing the steap gradients of the boundary layer.

\section{Conclusion and Outlook}

The assimilation of both velocity and temperature has been pursued for a case of cubic Rayleigh-B\'enard convection at $\mathrm{Ra}=10^6$ and $\mathrm{Pr}=0.7$. 
The results show that the PINN approach with a periodic activation function is advantageous for the assimilation in both directions compared to other widely used activation functions, which is primarily expressed by the high correlation coefficients of the inferred fields with the ground truth.
We attribute the advantages of a sine activation to it's ability to express the structures of the flow. An example of this is vortices, which are ubiquitous in turbulent flows and often appear as counter-rotating pairs or even clusters. To model just the multiple sign changes of a given velocity component of such structures, several monotonically activated neurons are required, whereas a single periodically activated neuron may be sufficient.
Furthermore, sine functions are resilient against the vanishing gradient problem (cf. \cite{Sitzmann2020}), since their derivative is close to zero only in small, periodically occurring regions. ELU or tanh, on the other hand, produce vanishing gradients for the negative tail or both tails of the respective functions.

This means that a periodic activation with sine functions eases the tension between computational speed, cost and accuracy, as it typically produces higher quality results than the other activation functions investigated, hyperbolic tangent and ELU, for the same network size.

However, the inferred fields still exhibit some deviations regarding the absolute values.
Further increasing the size of the MLP architecture mitigates this problem, as our results show decreasing errors for larger neural network layers.

The presented investigations also revealed a limitation regarding the benefits of a sine activation.
We found that, for temperature assimilation, sine activations were only beneficial for sufficiently large network sizes.

Overall, we found that the direction of assimilation, generating velocity fields from temperature data or vice versa, has further significant influences on the results:
Compared to velocity assimilation, temperature assimilation exhibited a critical range of the number of neurons of each hidden layer, which has to be exceeded to obtain viable results. Furthermore, the convergence of the training is shifted to a larger number of epochs for the temperature assimilation.
Due to the complexity of the task, there are several possible reasons for this: It could be caused by the different weighting of the PDEs in action, the overall conditioning of the PDE-terms that are crucial for the inference, or more mundane facts, e.g. that the MLPs have three fields to fit and two to infer in the temperature assimilation case, while the ratio is one to four for the velocity assimilation.
Since the temperature assimilation is more challenging, the deviations of the inferred fields from the ground truth are visually more prominent.
However, the convective heat flux statistics are predicted significantly better than for the velocity assimilation, as its regions prone to error have less influence on the statistics.

Future challenges include pushing for higher $\mathrm{Ra}$ and $\mathrm{Pr}$ numbers, for which it may not be feasible to provide fully resolved fields.
With respect to the necessary expressiveness for the temperature assimilation, $N_\mathrm{N}=128$ also proved to be close to the lower threshold for obtaining viable results, which we expect to increase for higher dimensionless numbers. 
At the same time, further features such as von Neumann boundary conditions or geometrically more complex boundary shapes should be implemented to improve the assimilation results by providing more information about the flow, which is typically available.
Besides that, network architecture, loss formulations, and the training process prove to be areas with a vast scope for design and thus also optimization potential. Relevant examples of such optimizations have been presented only recently by \cite{Toscano2024} and promise efficiency improvements.

\section*{Acknowledgment}
The authors gratefully acknowledge the scientific support and HPC resources provided by the German Aerospace Center (DLR). The HPC system CARA is partially funded by “Saxon State Ministry for Economic Affairs, Labour and Transport“ and „Federal Ministry for Economic Affairs and Climate Action”.

\bibliographystyle{plainnat}
\bibliography{references}

\appendix

\section{Hyperparameter study - number of hidden layers}
\label{sec:app_depth}

\begin{figure}[!h]
	\centering
	\includegraphics[width=0.99\linewidth]{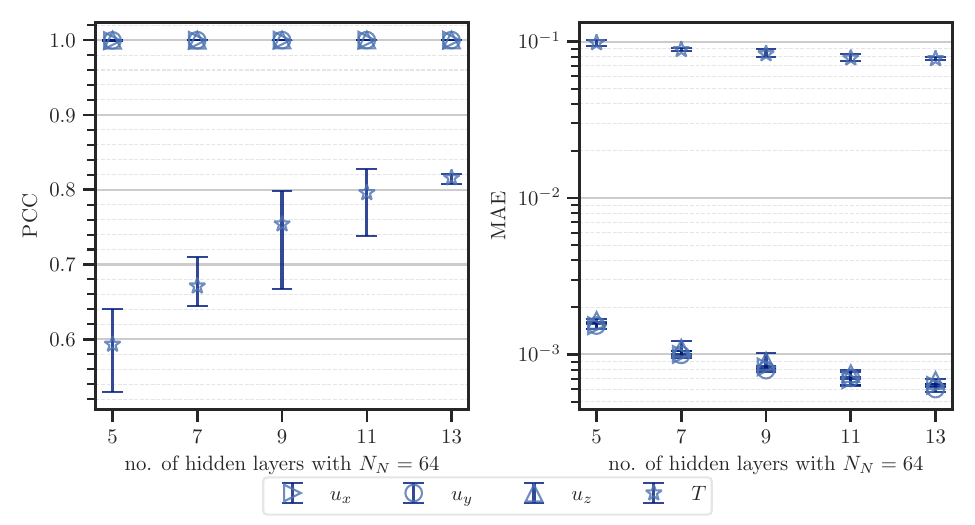}
	\caption{Correlation (left) and MAE (right) metrics of the velocity and temperature fields achieved after 5000 epochs of temperature-assimilation training for $N_\mathrm{N}=64$ with a varying number of hidden layers for the sine activation. The markers represent averages over multiple training runs and the error bars the respective minima and maxima.}
	\label{fig:Tassi_depth}
\end{figure}

While in the main part of the paper the layer width $N_N$ is varied to adjust the complexity of the neural network for different tasks and activation functions, this appendix aims to give an outlook on the variation of the number of hidden layers.
For this purpose, we test the temperature assimilation for sine activated PINNs with a constant layer width of $N_N=64$, which was below the threshold for obtaining usable results with five hidden layers of neurons.
Hence, three test runs each were performed for increasing numbers of hidden layers up to 13.
As expected, the results displayed in figure~\ref{fig:Tassi_depth} show an increase in assimilation performance for larger numbers of hidden layers.
Yet, the correlation values of a PINN of 5 hidden layers with $N_N=128$ are not reached by the configuration of 13 hidden layers with $N_N=64$.
Overall, the determination of an optimal aspect ratio for the MLP architecture requires further investigation, which should be part of future studies.

\section{Hyperparameter study - selected loss weights $\lambda_i$}
\label{sec:app_lambda}

In the section~\ref{sec:ML_opt} paragraph on weighting the loss components, two different sets of loss weights $\lambda_i$ are introduced. Specifically, there is one set for velocity assimilation ($\lambda_\mathrm{NSe}=10^{-2}, \lambda_\mathrm{CDe}=10^{-1}, \lambda_\mathrm{Coe}=\lambda_\mathrm{bounds}=\lambda_\mathrm{pc}=10^{-3}$) and one for temperature assimilation ($\lambda_\mathrm{NSe}=10^{-1}, \lambda_\mathrm{CDe}=10^{-2}, \lambda_\mathrm{Coe}=\lambda_\mathrm{bounds}=\lambda_\mathrm{pc}=10^{-3}$).
To underpin the distinction between a convection-diffusion driven velocity assimilation and a Navier-Stokes driven temperature assimilation, we also performed three training runs with the opposite set of weights.
The results in table~\ref{tab:weight_perf} show that distinguishing between the different assimilation mechanisms is indeed advantageous for the assimilation tasks.

\begin{table}[!bth]
	\centering
	\caption{Correlation Performance of the different assimilation tasks for a sine activated PINN with $N_N=128$ after 5000 epochs averaged over the respective number of runs (three except for the temperature assimilation with the corresponding weight set with five runs).}
	\begin{tabular}{c cccc|cccc}
		\hline\hline
		\multicolumn{1}{c}{} & \multicolumn{4}{c|}{\textbf{ $\boldsymbol{u}$ assimilation}} & \multicolumn{4}{c}{\textbf{$T$ assimilation}} \\
		\multicolumn{1}{c}{} & $PCC_{u_x}$ & $PCC_{u_y}$ & $PCC_{u_z}$ & $PCC_{T}$ & $PCC_{u_x}$ & $PCC_{u_y}$ & $PCC_{u_z}$ & $PCC_{T}$ \\
		\hline
		weight set $\boldsymbol{u}$ & \textbf{0.938} & \textbf{0.951} & \textbf{0.968} & 0.998 & 1.000 & 1.000 & 1.000 & 0.856 \\
		weight set $T$ & 0.892 & 0.886 & 0.929 & 0.999 & 1.000 & 1.000 & 1.000 & \textbf{0.946} \\
		\hline\hline
	\end{tabular}
	\label{tab:weight_perf}
\end{table}

\begin{figure}[!h]
	\centering
	\includegraphics[width=0.99\linewidth]{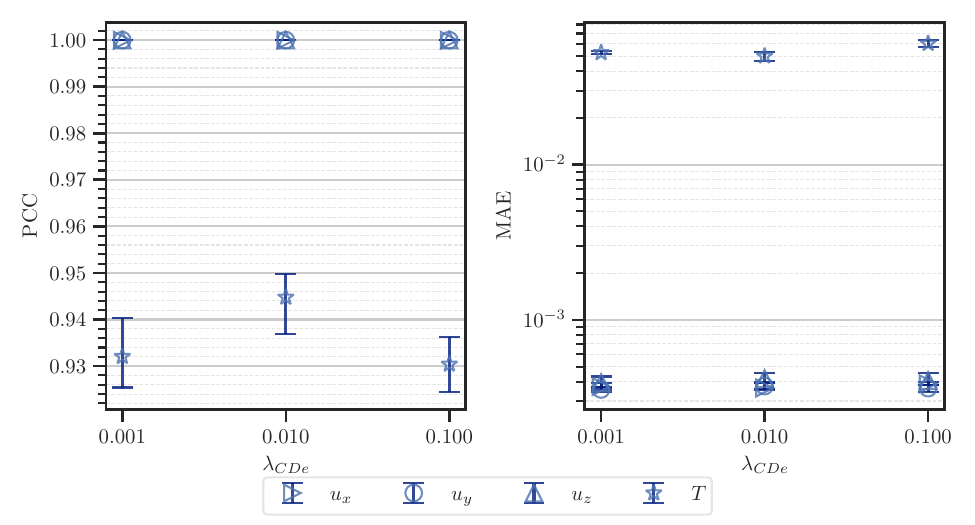}
	\caption{Correlation (left) and MAE (right) metrics of the velocity and temperature fields achieved after 5000 epochs of temperature-assimilation training for $N_\mathrm{N}=128$ with a varying convection-diffusion loss weight $\lambda_\mathrm{CDe}$ for the sine activation. The markers represent averages over multiple training runs and the error bars the respective minima and maxima.}
	\label{fig:Tassi_lambda}
\end{figure}

To provide more detail on how the convection-diffusion equation of the temperature affects the performance of the temperature assimilation, a small hyperparameter study of $\lambda_\mathrm{CDe}$ is provided by figure~\ref{fig:Tassi_lambda}.
For this study, the remaining set of weights is kept constant at $\lambda_\mathrm{NSe}=10^{-1}$ and $\lambda_\mathrm{Coe}=\lambda_\mathrm{bounds}=\lambda_\mathrm{pc}=10^{-3}$.
Besides the existing 5 runs for $\lambda_\mathrm{CDe}=10^{-2}$ from the main investigation, 3 additional runs were performed for $\lambda_\mathrm{CDe}=10^{-3}$ and $\lambda_\mathrm{CDe}=10^{-1}$.
Overall the final correlation coefficients and absolute errors achieved by the training exhibit only small variations, although the order of magnitude of $\lambda_\mathrm{CDe}$ was varied. 
This confirms, that the loss of the convection-diffusion equation is less relevant for the performance of the temperature assimilation than the Navier-Stokes loss.
Furthermore, the decrease in assimilation performance is slightly more pronounced for increased $\lambda_\mathrm{CDe}$.
A possible reason for this behavior could be the existence of a trivial solution $T = \mathrm{const.}$ for the convection-diffusion equation.
A similar behavior can be observed for the continuity equation with the trivial solution $\boldsymbol{u} = 0$ which leads to decaying PINN performances if weighted to strong (cf. \cite{Lucor2022}).

\section{Hyperparameter study - initialization factor $w$}
\label{sec:app_ini}

\begin{figure}[!h]
	\centering
	\includegraphics[width=0.99\linewidth]{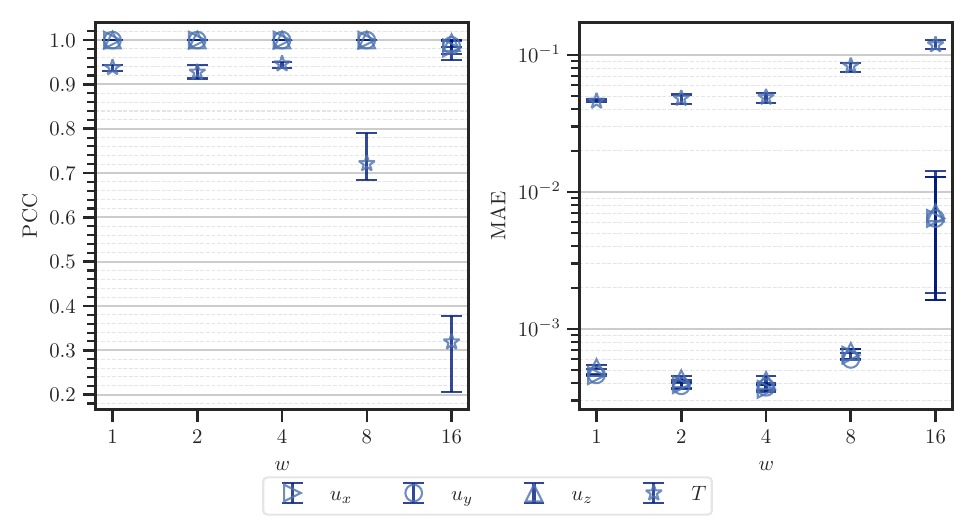}
	\caption{Correlation (left) and MAE (right) metrics of the velocity and temperature fields achieved after 5000 epochs of temperature-assimilation training  for $N_\mathrm{N}=128$ with a varying initialization parameter $w$ for the sine activation. The markers represent averages over multiple training runs and the error bars the respective minima and maxima.}
	\label{fig:Tassi_w1}
\end{figure}

As mentioned in section~\ref{sec:opti}, the sine activation also allows for different initialization strategies.
Here we varied the prefactor $w$ of the range of the random uniformly distributed initial weights of the first hidden layer.
Fig.~\ref{fig:Tassi_w1} shows the assimilation performance of a sine-activated PINN with $N_\mathrm{N}=128$ for $w \in \{1,2,4,8,16\}$ in the style of fig.~\ref{fig:Tassi_size1}, for which three training runs with different random initialization were carried out for each $w\neq4$.

It shows that the sine-activated PINN achieves the best correlation results of the inferred $T$ field and the lowest MAE of the fitted velocity fields for $w=4$.
However, the lowest MAE of the assimilated temperatures was obtained for $w=1$.
Overall, the performance of the PINN varies only slightly for $1\leq w \leq 4$, while it degrades significantly for higher values.
For $w=16$, even the correlation  coefficients of the fitted velocity components start to deteriorate.

A possible explanation for the poor performance of large $w$ is that they introduce unphysically high spatio-temporal wave numbers into the fields represented by the MLP. The high wave numbers are not regularized by the data loss terms, since a fit to data with lower wave numbers also works on the basis of the excessively high wave numbers.
This means that the wave numbers of neurons, that were initialized non-physically high, are not being reduced during training. This means that these neurons will either be suppressed by the physics losses during training, causing them to die, or they will persistently disturb the physics represented by the PINN.

Therefore, we would rather recommend the use of $w=1$, if a hyperparameter study of $w$ is not feasible, since its critical value will certainly dependent on the investigated flow.

\section{Pressure fields}
\label{sec:app_press}

\begin{figure}[!h]
	\centering
	\includegraphics[width=0.99\linewidth]{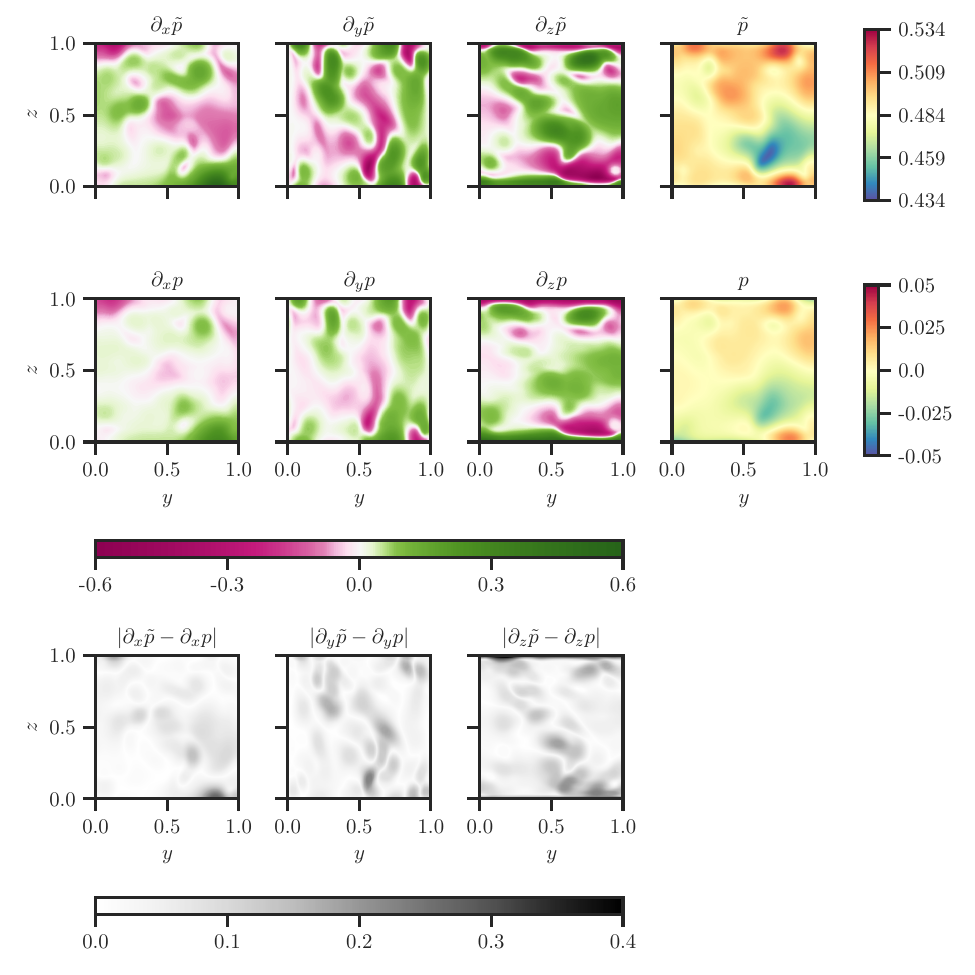}
	\caption{Comparison of the ground truth (top row) and PINN-generated (middle row) fields of the pressure and its spatial gradients in a central vertical cross-section for velocity assimilation with sine activation and $N_\mathrm{N}=128$.
	The bottom row comprises the contour plots of the gradients' absolute differences.}
	\label{fig:uassi_pfields}
\end{figure}

\paragraph{Velocity assimilation}
Fig.~\ref{fig:uassi_pfields} displays the comparison between the ground truth pressure field including its spatial gradients and the corresponding inferred fields.
It shows that the assimilated pressures are as good as the inferred velocity fields, since they exhibit the same characteristics:

All structures, such as the depression at $y\approx0.6, z\approx0.2$ and the rises near the boundaries, are represented. However, the PINN falls short in reconstructing the amplitudes.
This shows that, in this case, the velocities are mainly inferred by the convection diffusion equation and then the loss of the Navier-Stokes equations acts to reconstruct a pressure (gradient) field according to the inferred velocities.

\begin{figure}[!h]
	\centering
	\includegraphics[width=0.99\linewidth]{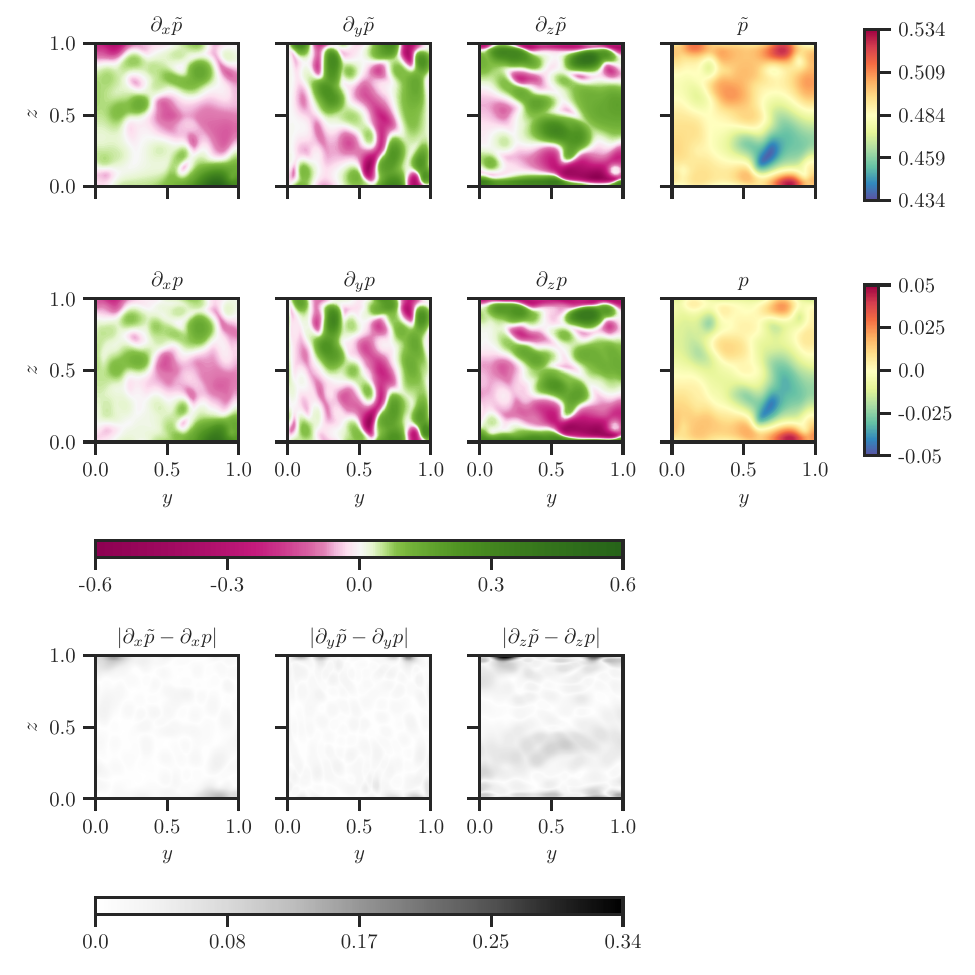}
	\caption{Comparison of the ground truth (top row) and PINN-generated (middle row) fields of the pressure and its spatial gradients in a central vertical cross-section for temperature assimilation with sine activation and $N_\mathrm{N}=128$.
	The bottom row comprises the contour plots of the gradients' absolute differences.}
	\label{fig:Tassi_pfields}
\end{figure}

\paragraph{Temperature assimilation}
The comparison of the pressure field and its gradients for the temperature assimilation case is displayed in fig.~\ref{fig:Tassi_pfields}.
Unlike velocity assimilation, in this case the pressure can be assimilated directly from the fitted velocity fields.
This leads to a better agreement of the inferred pressure fields with the ground truth than for velocity assimilation.
The largest remaining deviations occur for $\partial_z p$ in the same region as the largest bulk deviation of the inferred temperatures. 
This shows that the assimilation of both pressure and temperature poses a complication for the PINN approach.
This means that supporting one field with additional information, e.g. by implementing pressure taps in an experimental context, should improve the inference performance of both fields.

\end{document}